\documentstyle{elsart}
\begin{document}
\begin{frontmatter}
\title{QCD sum rules for $J/\psi$ in the nuclear medium: calculation of the
Wilson coefficients of gluon operators up to dimension 6}

\collab{Sungsik Kim and
Su Houng Lee}
\address{Department of Physics and Institute of Physics and Applied Physics, Yonsei University, Seoul 120-749, Korea}

\begin{abstract}

We calculate the Wilson coefficients of all dimension 6 gluon operators with
non zero spin in the correlation function between two heavy vector currents.
 For the twist-4 part,  we first  identify the three independent
gluon operators,  and then proceed with the calculation of the
Wilson coefficients using the fixed point gauge.   Together with the
previous calculation of the Wilson coefficients for the dimension 6 scalar
gluon operators
by Nikolaev and Radyushkin, our result completes the list of all the
Wilson coefficients of dimension 6 gluon operators in the correlation function
between heavy vector currents.   We apply our results to
investigate the mass  of $J/\psi$ in nuclear matter using QCD sum rules.
Using an upper bound estimate on the matrix elements of the dimension 6
gluon operators to linear order in density, we find that the density
dependent contribution from dimension 6 operators is  less than
$40\%$ of the dimension 4 operators with opposite sign.   The final result
gives about $-4$ MeV mass shift for the charmonium at rest in nuclear matter.

\end{abstract}

}

\end{frontmatter}

\section{Introduction}

Identifying higher twist operators and calculating their corresponding
Wilson coefficients are very important in several aspects.
 First, these provide a systematic building blocks to analyze the
available data from deep inelastic scattering(DIS) at lower $Q^2$ region.
Second, these contributions are essential for a realistic generalization of
QCD sum rule methods to finite baryon density.

The twist-4 operators were  classified and  its anomalous dimensions for
some of the operators were calculated  by S. Gottlieb\cite{Got78}.
However, the operators were over-determined and the independent set
of twist-4 operators appearing in the DIS were
first identified by Jaffe and Soldate\cite{Jaffe}, where the twist-4
operators  are of four quark type and quark gluon mixed type.
Among the uses of these results, the
available  DIS data were analyzed to determine the nucleon matrix elements
of the twist-4 operators\cite{Cho93,Lee94}.  These estimates were then used in
generalizing the QCD sum rule approaches for light vector mesons to finite
density\cite{HL92,Lee98}.

In the correlation function of two heavy vector currents,
only gluon  operators contribute in the operator product expansion(OPE).
This is so because in the heavy quark system, all the
heavy quark condensates are generated via gluonic
condensates\cite{Shi79,Rei85,Gen84}.
In dimension 6, there are scalar operators, twist-2
and twist-4 operators.
 For the scalar gluonic operators at dimension 6, there are two
independent operators.  In ref.\cite{Nik83}, the two
were identified and the corresponding Wilson coefficient were calculated .
  For twist-2 gluon operator, the calculation for the leading order(LO)
Wilson coefficient is simple and its matrix element is just the second
moment of the gluon structure function.

The twist-4 dimension 6 gluon operators are more involved.  In this work,
we have identified  the three independent local gluon operators and
calculated  their corresponding  LO Wilson coefficients in the
correlation function between two heavy vector currents.
This result is new and complimentary to a previous work\cite{BB99}
on gluon twist-4 operators, where they start
from certain diagrams and identify the three independent twist-4 gluon
structure  functions.
Together with the
previous calculation of the Wilson coefficients for the dimension 6 scalar
gluon operators
by Nikolaev and Radyushkin\cite{Nik83}, our result completes the list of all the
Wilson coefficients of dimension 6 gluon operators in the correlation function
between heavy vector currents.
As an application,
we will use our result in QCD sum rule approach to
investigate the property of  $J/\psi$ in nuclear matter.

This is particularly interesting because the
on-going discussion of $J/\psi$ suppression in RHIC as a possible signal for
quark gluon plasma\cite{Matsui86}, inevitably requires a detailed knowledge of the changes of
 $J/\psi$ properties in ``normal" nuclear matter\cite{Vogt99}.
 Furthermore, the large charm quark mass $m_c \gg \Lambda_{QCD}$ provides
a natural renormalization point for which a perturbative QCD
expansion is partly possible.  In fact, studies have shown that
the   multi-gluon
exchange  between a $ c \bar{c}$ pair and nucleons might induce a bound
 $ c \bar{c}$ state with even light nuclei\cite{Bro90,Was91,Luk92,Kaid92,Dim96,Bro97,Ter98}.
In such analysis,  the low energy multi-gluon potential was
modeled either from the effective theory obtained in the
infinitely large $m_Q$ limit\cite{Luk92,Kaid92,Dim96}  or  from extrapolating
the high energy scattering via pomeron exchange to lower
energy\cite{Bro90}.
Although both approaches, gave similar binding for the $J/\psi$ in
nuclear matter, it is not clear how reliable these results are unless
one systematically calculates the corrections.

In order to confirm this findings in an alternative but a systematic
approach and to establish  a basis for further studies,
 we have previously applied the QCD sum rules\cite{Shi79,Rei85} to heavy
quark system in nuclear medium and calculated the mass of
 $J/\psi$ and $\eta_c$ in nuclear medium\cite{Kli99,Hay99}.  This was the generalization
of the sum rule method for the light vector mesons in medium\cite{HL92}
to the  heavy quark system.   It was found
that the mass of $J/\psi$ ($\eta_c$) would reduce by
  7 MeV (5 MeV), which is indeed consistent with previous
results based on completely different methods\cite{Was91,Luk92,Bro97,Ter98}.
 However, it was not possible to reliably estimate the uncertainties
of the result, because the contribution from the
operator product expansion was truncated at the leading
 dimension 4 operators.
To overcome the limitations , we will here make use of our calculation to
include the complete dimension 6 contributions.
Unfortunately, at present, there is no data to identify the nucleon
expectation value of any of the dimension 6 operators(scalar, twist-2,
twist-4) nor is there any lattice result.
Nevertheless, in this
article, we will estimate the nucleon matrix elements of these
operators  and then use  QCD sum rule approach to study the reliability
of our previous result on
the mass shift of $J/\psi$ in nuclear matter.
If in the future,  the matrix elements are better known,
we  will be able to  study
the non trivial momentum dependence of the $J/\psi$ in
nuclear medium, as has been done for the light vector
mesons\cite{Lee98}.   The momentum dependence is especially interesting
because, there are inelastic channels
opening when the  charmonium system is moving with respect to the
medium.   For example, when the charmonium is moving with sufficient
velocity, it will have enough energy to interact with a nucleon to
produce a D meson and a charmed nucleon.
This effect would be very important in relation to
 $J/\psi$ suppression in RHIC.

In section 2 we characterize all gluon operator up to dimension 6
and obtain identities to be used to reduce the operators to an
independent set.
In section 3, we calculate the Wilson coefficients for the independent
set of gluon operators and show current conservation.
In section 4, we calculate the moments and perform a moment
sum rules analysis to calculate the $J/\psi$ mass in nuclear matter.
We conclude with some discussions.   The appendix includes some
detailed calculation of the Wilson coefficients.

\section{Operators}

In the operator product expansion of heavy quark system,
only gluonic condensates are relevant.
This is so because all the heavy quark condensates can be
related to the gluon condensates via heavy quark
expansion~\cite{Shi79,Rei85,Gen84}.
This is also true in nuclear medium since there are no valence
charm quarks to leading order in density and any interaction with
the medium is gluonic.  Let us start by categorizing gluonic
operators up to dimension 6, which does not vanish in nuclear
matter.

For dimension 4 operators, the scalar and twist-2 gluon operators contribute
\cite{Kli99,Hay99},
\begin{eqnarray}
g^2G^a_{\mu\nu}G^a_{\mu\nu},~~~~g^2G^a_{\mu\alpha}G^a_{\nu \alpha}.
\end{eqnarray}

For dimension 6 operators,  one can think of generating a number
of gluon operators constructed from three gluon fields $G^a_{\mu\nu}$
or two gluon fields with two covariant derivatives.  However,
for scalar operators, there are only two independent
scalar operators\cite{Nik83}.  They are,
\begin{eqnarray}
g^3f^{abc}G^a_{\mu\nu}G^b_{\mu\alpha}G^c_{\nu\alpha},~~~~~
g^2G^a_{\mu\alpha}G^a_{\nu\alpha;\nu\mu}.
\end{eqnarray}
The second operator can also be written in terms of four quark operator using the
equation of motion
\begin{eqnarray}
\label{em}
G^a_{\mu\nu;\nu}=g \bar{q} \gamma_\mu \frac{\lambda^a}{2} q =gj_\mu^a,
\end{eqnarray}

As for the spin 2 operators in dimension 6, which are also called dimension 6 twist 4
(twist=dimension-spin) operator,  we can first
categorize possible operators as follows.  First, there is again
one operator with three gluon field strength tensor.  Then,
assuming the
free symmetric and traceless indices to be $\mu$ and $\nu$,  depending on whether
or not the free index goes into the covariant derivative,
there are 6 more operators possible.    Starting with the
three gluon operator, the 7 are,
\begin{eqnarray}
\label{gluonop}
&&g^3 f^{abc} G^a_{\mu\kappa} G^b_{\nu\lambda} G^c_{\kappa\lambda}
\equiv g^3 f G^3_{\mu\nu}\nonumber\\[20pt]
&&g^2 G^a_{\kappa\lambda} G^a_{\kappa\lambda;\mu\nu} \nonumber\\
&&g^2 G^a_{\mu\kappa} G^a_{\nu\lambda;\kappa\lambda},\,\,
 g^2 G^a_{\mu\kappa} G^a_{\nu\kappa;\lambda\lambda},\,\,
 g^2 G^a_{\mu\kappa} G^a_{\nu\lambda;\lambda \kappa} \nonumber\\
&&g^2 G^a_{\mu\kappa} G^a_{\kappa\lambda;\lambda\nu},\,\,
     g^2 G^a_{\kappa\lambda} G^a_{\mu\kappa;\lambda\nu}.
\end{eqnarray}

However, they are not independent.  Using the identities in
appendix \ref{identity}, one can show that there are three independent spin-2
operators at dimension 6.
The set we will use are,
\begin{eqnarray}
\label{ope4}
    g^2\,G^a_{\kappa\lambda} G^a_{\kappa\lambda;\mu\nu},~~
    g^2\,G^a_{\mu\kappa} G^a_{\nu\lambda;\lambda\kappa},~~
    g^2\,G^a_{\mu\kappa} G^a_{\kappa\lambda;\lambda\nu}.
\end{eqnarray}

It is interesting to note that in reference \cite{BB99} starting from
certain diagrams with two, three and four gluon exchange in the
t-channel, they were able to derive three twist-4 gluon
distribution amplitudes, from which one can calculate the nucleon
matrix element of the three independent operators.

As for the spin 4 dimension 6 operator, it is just the twist-2
gluon operator.

\section{Polarization (OPE)}

Having established the independent gluon operators in dimension 6,
we will calculate their LO Wilson coefficients in the correlation function
between two vector currents made of heavy quarks,
$j_\mu={\bar h}\gamma_\mu h$.

\begin{eqnarray}
\label{polarization1}
\Pi_{\mu\nu}(q)&=&i\int d^4 x e^{iqx} \langle  {\rm T} \{ j_\mu(x) j_\nu(0) \}\rangle_\rho
\nonumber\\
&=&i\int d^4 x e^{iqx}
\langle  {\rm Tr} \left[ \gamma_\mu S(x,0) \gamma_\nu S(0,x)) \right] \rangle_\rho
\nonumber\\
&=&i\int \frac{d^4 k}{(2\pi)^4}
\langle \, {\rm Tr}[\, \gamma_\mu S(k+q) \gamma_\nu {\tilde S}(k)]\,\rangle_\rho,
\end{eqnarray}

where $\langle \cdot \rangle_\rho$ represents the expectation value
at finite nuclear density $\rho$.
The fourier transforms are defined  by
\begin{eqnarray}
&&iS(p)=\int d^4x e^{ipx}\,iS(x,0) \nonumber\\
&&i{\tilde S}(p)=\int d^4x e^{-ipx}\,iS(0,x).
\end{eqnarray}
$S(x,0)$ is the heavy quark propagator in the presence of
external gluon field \cite{Niv84}.
To calculate the Wilson coefficients, we obtain  the
quark propagator in the presence of external gauge fields
in the  Fock-Schwinger gauge~\cite{Nik83,Niv84},
\begin{eqnarray}
x_\mu A^a_\mu (x)=0.
\end{eqnarray}
In  appendix \ref{propagator} we list the momentum space representation
of the quark propagator multiplying the gauge invariant
gluon operators.
We will use these quark propagators in eq.(\ref{polarization1})
and extract the Wilson coefficients by collecting the
appropriate tensor and gluon structures.

In general the polarization tensor in eq.(\ref{polarization1})
will have two invariant functions.  They can be divided into the longitudinal and
transverse part of the external three momentum, which are the
following when the nuclear matter is at rest.
\begin{eqnarray}
\label{twopol}
\Pi_L=\frac{1}{k^2} \Pi_{00},~~~~\Pi_T
=-\frac{1}{2}(\frac{1}{q^2}\Pi^\mu_\mu+\frac{1}{k^2}\Pi_{00}),
\end{eqnarray}
where
$q=(\omega,k)$ is the external momentum.

In the vacuum or in the limit when $k\rightarrow0$ they become
the same so that there is only one invariant function
\begin{eqnarray}
\label{polarization2}
\Pi_L(\omega^2)=\Pi_T(\omega^2)=\frac{-1}{3\omega^2}
g^{\mu\nu}\Pi_{\mu\nu}\equiv \Pi(\omega^2).
\end{eqnarray}
So in this work, we will construct the sum rule for $\Pi(\omega^2)$
at $k\rightarrow 0$ and nuclear matter at rest.
However, in the calculation of the OPE, we will start from the
general expression of eq.(\ref{polarization1}) at nonzero value of $k$
and calculate each tensor structure separately.
This way, current conservation will be a nontrivial check of our calculation and
the generalization to  $k\neq 0$ will be straightforward\cite{SL00}.

In the following subsections, we will summarize the OPE
for operators of dimension 4 and  6.
The new results of the present work are the OPE for
dimension 6 spin 2 and  dimension 6 spin 4 operators.

\subsection{scalar contributions}

Here we summarize the OPE of scalar operators of dimension 4 and
dimension 6 used in the vacuum sum rule\cite{Nik83}.

\begin{eqnarray}
\label{scalarope}
\Pi^{\rm scalar}_{\mu\nu} (q)&=&(q_\mu q_\nu-g_{\mu\nu}q^2)\Big[
C^{pert.} +
C^0_{G^2} \langle g^2 G^a_{\alpha\beta} G^a_{\alpha\beta} \rangle \nonumber\\&&+
C^0_{G^3} \langle g^3 f^{abc} G^a_{\alpha\beta} G^b_{\beta\lambda} G^c_{\lambda\alpha} \rangle
+C^0_{j^2} \langle g^4 j^a_\kappa j^a_\kappa \rangle\Big],
\end{eqnarray}
where
{\scriptsize
\begin{eqnarray}
&&C^0_{G^2}=\frac{1}{48\pi^2 (Q^2)^2}[-1+3J_2-2J_3],
\nonumber\\
&&C^0_{G^3}=\frac{1}{72\pi^2 (Q^2)^3}\left[ \frac{2}{15}-\frac{1}{10}y +4J_2-\frac{31}{3}J_3+
\frac{43}{5}J_4-\frac{12}{5}J_5 \right],
\nonumber\\
&&C^0_{j^2}=\frac{1}{36\pi^2 (Q^2)^3}
\left[ \frac{41}{45}-\frac{3}{5}y+(\frac{2}{3}+\frac{1}{3}y)J_1-J_2-\frac{4}{9}J_3-
\frac{26}{15}J_4+\frac{8}{5}J_5 \right],
\end{eqnarray}}
and
\begin{eqnarray}
\label{defJ}
J_N(y)=\int^1_0 \frac{dx}{\left[ 1+x(1-x) y \right]^N},
\end{eqnarray}
with $y=Q^2/m^2$
and $m$ being the heavy quark mass.
This will give the following contribution to $\Pi (\omega^2)$ defined in
eq.(\ref{polarization2}).

\begin{eqnarray}
\Pi (\omega^2) & =& C^{pert.} +
C^0_{G^2} \langle g^2 G^a_{\alpha\beta} G^a_{\alpha\beta} \rangle +
C^0_{G^3} \langle g^3 f^{abc} G^a_{\alpha\beta} G^b_{\beta\lambda} G^c_{\lambda\alpha} \rangle
\nonumber \\
& & +C^0_{j^2} \langle g^4 j^a_\kappa j^a_\kappa \rangle.
\end{eqnarray}

For later convenience, we will use
$g^3 f^{abc} G^a_{\mu\nu} G^b_{\nu\lambda} G^c_{\lambda\mu} =
\frac{1}{2} g^2 G^a_{\mu\nu}G^a_{\mu\nu;\alpha\alpha}+g^4 j^a_\mu j^a_\mu $
and rewrite the OPE as follows,

\begin{eqnarray}
\Pi (\omega^2)
 & =& C^{pert.} +
C_{G^2} \langle \frac{\alpha_s}{\pi} G^a_{\mu\nu} G^a_{\mu\nu} \rangle +
C_{GD^2G} \langle \frac{\alpha_s}{\pi} G^a_{\mu\nu} G^a_{\mu\nu;\kappa\kappa} \rangle
\nonumber \\
&& +C_{j^2}\cdot
\frac{2}{3} \langle \frac{\alpha_s}{\pi} G^a_{\alpha\kappa} G^a_{\alpha\lambda;\lambda\kappa}
\rangle ,
\end{eqnarray}
where
\begin{eqnarray}
C_{G^2} & = &\frac{1}{12 (Q^2)^2}[-1+3J_2-2J_3], \nonumber\\
C_{GD^2G} & = & \frac{1}{1080 (Q^2)^3}[4-3y +120J_2-310J_3+258J_4 -72J_5], \nonumber\\
C_{j^2} & = & \frac{1}{1080 (Q^2)^3}[(4-3y +120J_2-310J_3+258J_4 -72J_5)\nonumber\\
&&-180+120y-(120+60y)J_1-300J_2+1320J_3-720J_4 ].
\end{eqnarray}
Note $Q^2=-\omega^2$ in our kinematical limit.

\subsection{dimension 4 spin 2 operator}

This part has been calculated  recently~\cite{Kli99,Hay99}.
Here we summarize the results.

\begin{eqnarray}
\label{dim4s2}
\Pi^{4,2}_{\mu\nu}(q) & = & \frac{1}{Q^2}
\bigg[{\cal I}^2_{\mu\nu}+\frac{1}{Q^2} (q_\rho q_\mu {\cal I}^2_{\rho\nu}+
q_\rho q_\nu {\cal I}^2_{\rho\mu})
\nonumber \\
&&+g_{\mu\nu} \frac{q_\rho q_\sigma}{Q^2} {\cal J}^2_{\rho\sigma}+
\frac{q_\mu q_\nu q_\rho q_\sigma}{Q^4}
({\cal I}^2_{\rho\sigma}+{\cal J}^2_{\rho\sigma})\bigg],
\end{eqnarray}
where
\begin{eqnarray}
\label{IJ42}
{\cal I}^2_{\mu\nu}&=&\langle\frac{\alpha_s}{\pi} G^a_{\sigma\mu}G^a_{\sigma \nu}\rangle
\left[ \frac{1}{2}+(1-\frac{1}{3}y)J_1-\frac{3}{2}J_2\right]
\nonumber\\[5pt]
{\cal J}^2_{\mu\nu}&=&\langle\frac{\alpha_s}{\pi} G^a_{\sigma\mu}G^a_{\sigma \nu}\rangle
\left[ -\frac{7}{6}+(1+\frac{1}{3}y)J_1-\frac{1}{2}J_2+\frac{2}{3}J_3 \right].
\end{eqnarray}

\subsection{contribution from dimension 6 and spin 4}

\begin{figure}[htb]
\vbox to 3.0in{\vss
   \hbox to 2.5in{\includegraphics{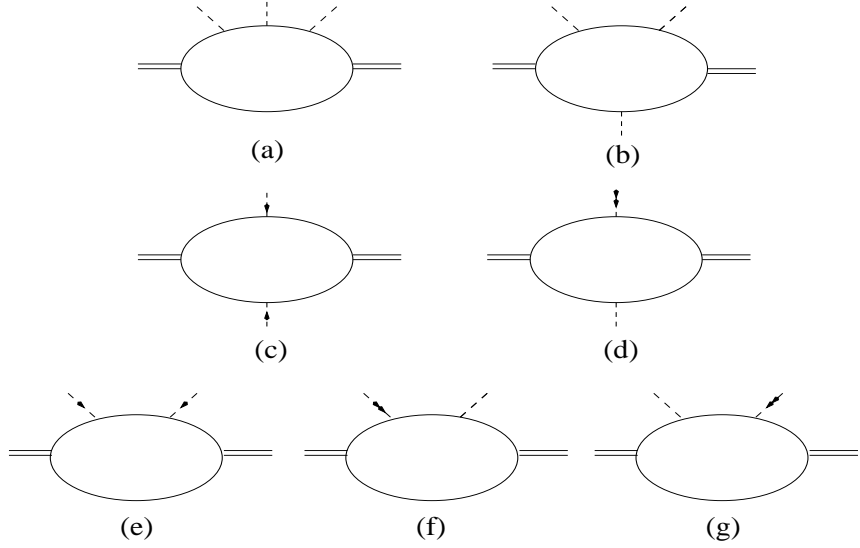}\hss}}
\label{fig:diagrams}
\caption{1-loop diagrams describing interactions with the gluonic field
giving dimension 6 operators: The dashed line represents an external gluon field $G$
and the dashed line
with $n$ arrows represents an external gluon field with $n$ derivatives
$D^nG$.\label{fig:feynman}}
\end{figure}

The diagrams describing interactions with the gluonic field in dimension 6
are shown in Fig \ref{fig:feynman}.

The calculation for this part is straightforward.  We substitute into
eq.(\ref{polarization1}) parts of the quark propagator containing $D$'s and $G$'s
 (summarized in appendix C) so that when these from the two propagators are
combined, yield terms proportional to $GDDG$.
In Fig.\ref{fig:feynman}, these come from Fig.\ref{fig:feynman}(c)
to Fig.\ref{fig:feynman}(g).
Then we extract the traceless and symmetric spin-4 part of the operator $GDDG$.
The final result for the dimension 6 spin 4 part of the operators yields,
\begin{eqnarray}
\label{dim6s4}
\Pi^{6,4}_{\mu\nu}(q) & = &
\frac{q_\kappa q_\lambda}{(Q^2)^3}
\bigg[I^4_{\kappa\lambda\mu\nu}+
\frac{1}{Q^2} (q_\rho q_\mu I^4_{\kappa\lambda\rho\nu}+
q_\rho q_\nu I^4_{\kappa\lambda\rho\mu}) \nonumber \\
&& +g_{\mu\nu} \frac{q_\rho q_\sigma}{Q^2} J^4_{\kappa\lambda\rho\sigma}+
\frac{q_\mu q_\nu q_\rho q_\sigma}{Q^4}
(I^4_{\kappa\lambda\rho\sigma}+J^4_{\kappa\lambda\rho\sigma})
\bigg],
\end{eqnarray}
where
\begin{eqnarray}
\label{IJ64}
I^4_{\mu\nu\rho\sigma} & = &
\left[-\frac{266}{45} +\left(-\frac{20}{3} +\frac{22}{15} y \right)J_1
+\frac{138}{5} J_2 -\frac{916}{45} J_3+\frac{16}{3} J_4
\right]
\nonumber\\
&&\times\langle
\frac{\alpha_s}{\pi}G^a_{\rho\kappa} G^a_{\sigma\kappa;\mu\nu}\rangle
\nonumber\\[10pt]
J^4_{\mu\nu\rho\sigma}& =& \left[
\frac{362}{45}-\left( 4 +\frac{22}{15}y\right) J_1
-\frac{94}{15} J_2 -\frac{44}{45} J_3 +\frac{16}{3} J_4-\frac{32}{15} J_5
\right]
\nonumber\\
&&\times\langle
\frac{\alpha_s}{\pi}
G^a_{\rho\kappa} G^a_{\sigma\kappa;\mu\nu}\rangle.
\end{eqnarray}

\subsection{contributions from dimension 6 and spin 2}

The calculation for this part goes in a similar way as in the
previous subsection.  However, this time all the graphs in Fig.1
contribute.  Moreover, we have to include all possible combination
of $G$ and $D$'s that makes up dimension 6.  We then extract the
spin 2 part of each operator and use the identities given in the
appendix to reduce the operators into the independent three
in eq.(\ref{ope4}).
An example needed to extract the spin 2 part from a general operator
is given in appendix D.
Finally we find,
\begin{eqnarray}
\label{dim6s2}
\Pi^{6,2}_{\mu\nu}(q) & = &
\frac{1}{(Q^2)^2}
\bigg[I^2_{\mu\nu}+\frac{1}{Q^2} (q_\rho q_\mu I^2_{\rho\nu}+
q_\rho q_\nu I^2_{\rho\mu}) \nonumber \\
&&
+g_{\mu\nu} \frac{q_\rho q_\sigma}{Q^2} J^2_{\rho\sigma}+
\frac{q_\mu q_\nu q_\rho q_\sigma}{Q^4}
(I^2_{\rho\sigma}+J^2_{\rho\sigma})
\bigg],
\end{eqnarray}
where
{\scriptsize
\begin{eqnarray}
\label{IJ62}
I^2_{\mu\nu}&=&
\langle\frac{\alpha_s}{\pi} G^a_{\kappa\lambda} G^a_{\kappa\lambda;\mu\nu} \rangle
\left[\frac{31}{240}-\frac{1}{60}y+\left(\frac{13}{24}+\frac{1}{48}y\right)J_1
-\frac{115}{48}J_2+\frac{21}{8}J_3-\frac{9}{10}J_4
\right]\nonumber\\
&&+\langle\frac{\alpha_s}{\pi} G^a_{\mu\kappa} G^a_{\nu \lambda;\lambda\kappa}
\rangle
\left[ -\frac{739}{720}+\frac{2}{15}y+\left(-\frac{9}{8}+\frac{3}{16}y\right)J_1
+\frac{133}{48}J_2+\frac{1}{72}J_3-\frac{19}{30}J_4
\right]\nonumber\\
&&+\langle\frac{\alpha_s}{\pi} G^a_{\mu\kappa} G^a_{\kappa\lambda;\lambda\nu}
\rangle
\left[ \frac{293}{240}-\frac{3}{10}y+\left(\frac{55}{24}+\frac{1}{16}y\right)J_1
-\frac{131}{16}J_2+\frac{145}{24}J_3
-\frac{41}{30}J_4
\right]\nonumber\\[12pt]
J^2_{\mu\nu}&=&
\langle\frac{\alpha_s}{\pi} G^a_{\kappa\lambda} G^a_{\kappa\lambda;\mu\nu}
\rangle
\left[\frac{103}{240}+\frac{1}{60}y+\left(\frac{5}{24}-\frac{7}{48}y\right)J_1
-\frac{59}{48}J_2+\frac{31}{24}J_3
-\frac{11}{10}J_4+\frac{2}{5}J_5
\right]\nonumber\\
&&+\langle\frac{\alpha_s}{\pi} G^a_{\mu\kappa} G^a_{\nu\lambda;\lambda\kappa}
\rangle
\left[\frac{71}{240}-\frac{2}{15}y+\left(-\frac{1}{8}+\frac{1}{48}y\right)J_1
+\frac{61}{48}J_2-\frac{61}{24}J_3+\frac{29}{30}J_4
+\frac{2}{15}J_5
\right]\nonumber\\
&&+\langle\frac{\alpha_s}{\pi} G^a_{\mu\kappa} G^a_{\kappa\lambda;\lambda\nu}
\rangle
\left[\frac{29}{240}+\frac{3}{10}y-\left(\frac{1}{24}+\frac{7}{16}y\right)J_1
+\frac{31}{48}J_2-\frac{23}{24}J_3+\frac{11}{30}J_4
-\frac{2}{15}J_5
\right].
\end{eqnarray}}

There are few checks we can perform to verify our calculation.
First,
although the polarization function in eq.(\ref{polarization1})
has only two invariant  parts,  we performed the calculation
directly leaving the indices $\mu, \nu$ free.  This means that we
have obtained and calculated the 6 different possible tensor structure,
given in appendix B, separately.  Therefore,  showing current conservation,
 $q^\mu \Pi_{\mu\nu}=q^\nu \Pi_{\mu\nu}=0$ is a non-trivial check.
 Another indirect check is that the final result is regular
at $Q^2=0$.  That is, making a small $Q^2$ expansion
one can show that all
 the Wilson coefficients in eq.(\ref{IJ42}), eq.(\ref{IJ64}) and
eq.(\ref{IJ62}) are regular.

Eq.(\ref{scalarope}), eq.(\ref{dim4s2}), eq.(\ref{dim6s4}) and eq.(\ref{dim6s2})
form the complete OPE up to dimension 6 operators.  Since QCD is renormalizable,
there are no other power correction up to this dimension and the Wilson coefficients
will be an asymptotic expansion in $1/\log Q^2$.    In the vacuum, only scalar
operators will contribute.  However, in medium or when the expectation value is
with respect to a nucleon state, the tensor operators will also contribute.
As an application of our result, we will apply our OPE to the QCD sum rule
analysis for the $J/\psi$ in nuclear matter.

\subsection{sum of tensor parts}

As we discussed before we will take the nuclear matter to be at rest.
Moreover, we will use linear density approximation to evaluate
the matrix elements,
\begin{eqnarray}
\label{linear}
\langle  \cdot \rangle_\rho=\langle  \cdot  \rangle_0
+\frac{\rho}{2m_N} \langle N|\cdot |N \rangle
\end{eqnarray}
where $\langle \cdot \rangle_0$ is the vacuum expectation value,
$\rho$ is the nuclear density, $m_N$ the nucleon mass, and
the nucleon state $|N\rangle$ is normalized as
 $\langle N|N \rangle= (2\pi)^3 2\omega_N \delta^3(0)$.
 Then, the OPE  from dimension 4 and
dimension 6 operators in eq.(\ref{dim4s2}), eq.(\ref{dim6s4}) and
eq.(\ref{dim6s2}) give the following contribution to $\Pi$.
\begin{eqnarray}
\Pi (\omega^2)
 & =& \Pi_{4,2} (\omega^2)+\Pi_{6,2} (\omega^2)+\Pi_{6,4} (\omega^2)
\nonumber\\
&=& \frac{\rho}{2m_N} \left[C_{G_2}G_2 +(C_X X+C_Y Y+C_Z Z)
+C_{G_4} G_4 \right],
\end{eqnarray}
where
{\scriptsize
\begin{eqnarray}
C_{G_2}&=&\frac{m_N^2}{12 (Q^2)^2}[9-(12+2y)J_1+9J_2-6J_3], \nonumber\\
C_X&=&\frac{m_N^2}{240 (Q^2)^3}[-85-2y+(-70+25y)J_1+365J_2-390J_3+252J_4-72J_5],\nonumber\\
C_Y&=&\frac{m_N^2}{720 (Q^2)^3}[25+48y+(270-45y)J_1-1185J_2+1370J_3-408J_4-72J_5],\nonumber\\
C_Z&=&\frac{m_N^2}{240 (Q^2)^3}[-95-36y+(-130+75y)J_1+375J_2-190J_3+16J_4+24J_5],\nonumber\\
C_{G_4}&=&\frac{m_N^4}{108 (Q^2)^3}[205-(210+33y)J_1+99J_2-262J_3+240J_4-72J_5],
\end{eqnarray}}
where again in our kinematical limit $Q^2 =-\omega^2$
and the scalar parts of the matrix elements in nuclear matter
to leading density come from the following nucleon expectation
values,
\begin{eqnarray}
\label{opall}
\langle N|\frac{\alpha_s}{\pi} G^a_{\sigma\mu} G^a_{\sigma\nu} |N \rangle
& =& G_2\,(p_\mu p_\nu -\frac{1}{4} m_N^2 g_{\mu\nu}),
\nonumber\\
\langle N| \frac{\alpha_s}{\pi} G^a_{\kappa\lambda} G^a_{\kappa\lambda;\mu\nu} |N \rangle
& =& X\,(p_\mu p_\nu -\frac{1}{4} m_N^2 g_{\mu\nu}),
\nonumber\\
\langle N| \frac{\alpha_s}{\pi} G^a_{\mu\kappa} G^a_{\nu\lambda;\lambda\kappa} |N \rangle
&=& Y\,(p_\mu p_\nu -\frac{1}{4} m_N^2 g_{\mu\nu}),
\nonumber\\
\langle N| \frac{\alpha_s}{\pi} G^a_{\mu\kappa} G^a_{\kappa\lambda;\lambda\nu} |N \rangle
& =& Z\,(p_\mu p_\nu -\frac{1}{4} m_N^2 g_{\mu\nu}),
\nonumber\\
\langle N| \frac{\alpha_s}{\pi} G^a_{\mu\kappa} G^a_{\nu\kappa;\alpha\beta} |N \rangle
&=& G_4\,\bigg[p_\mu p_\nu p_\alpha p_\beta +\frac{m_N^4}{48}(g_{\mu\nu} g_{\alpha\beta}+
g_{\mu\alpha} g_{\nu\beta} +g_{\mu\beta} g_{\nu\alpha})
\nonumber\\
&&-\frac{1}{8} m_N^2 (p_\mu p_\nu g_{\alpha\beta}
+p_\mu p_\alpha g_{\nu\beta}+p_\mu p_\beta g_{\alpha\nu}
\nonumber\\
&&+p_\nu p_\alpha g_{\mu\beta}+p_\nu p_\beta g_{\mu\alpha}
+p_\alpha p_\beta g_{\mu\nu})
\bigg].\nonumber\\
\end{eqnarray}
Note that here we chose the nucleon four momentum
to be $p=(m_N,0,0,0)$.
We discuss the magnitudes of these nucleon matrix elements in
section \ref{moment}.

\section{Moment sum rule for $J/\psi$ at rest\label{moment}}
\subsection{moments}

As an application of our result, we will use our result to refine our previous
work to calculate  the mass shift of the $J/\psi$ in nuclear medium\cite{Kli99}
using the moment sum rule \cite{Shi79,Rei81}.
The moments of the
polarization function is defined as,
\begin{eqnarray}
M_n(Q_0^2)= \frac{1}{n!} \left( -\frac{d}{dQ^2}\right)^n \Pi(Q^2)|_{Q^2=Q_0^2}.
\end{eqnarray}
where in our kinematics, $Q^2=-\omega^2$.
Direct evaluation of these moments using the OPE gives, up to
dimension 4~\cite{Rei81,Kli99},
\begin{eqnarray}
M_n(\xi)=A_n^V(\xi)\left[ 1+a_n(\xi) \alpha_s +b_n(\xi) \phi^4_b+c_n(\xi) \phi^4_c \right],
\end{eqnarray}
where
\begin{eqnarray}
\label{phibc}
\phi^4_b&=&\frac{4\pi^2}{9}\frac{\langle \frac{\alpha_s}{\pi}G^2\rangle}{(4m_c^2)^2}
\nonumber\\
\phi^4_c&=&\frac{2\pi^2}{3}\frac{G_2}{(4m_c^2)^2} m_N \rho_N
\label{eqn:4}
\end{eqnarray}
and
\begin{eqnarray}
A_n^V(\xi)&=& \frac{3}{4\pi^2} \frac{2^n (n+1)(n-1)!}{(2n+3)!!} (4m^2)^{-n}
(1+\xi)^{-n}\, _2F_1(n,\frac{1}{2},n+\frac{5}{2};\rho)
\nonumber\\
b_n(\xi)&=& -\frac{n(n+1)(n+2)(n+3)}{2n+5}(1+\xi)^{-2}
\frac{_2F_1(n+2,-\frac{1}{2},n+\frac{7}{2};\rho)}
{_2F_1(n,\frac{1}{2},n+\frac{5}{2};\rho)}
\nonumber\\
c_n(\xi)&=& b_n(\xi)-\frac{4n(n+1)}{3(2n+5)(1+\xi)^2}
\frac{_2F_1(n+1,\frac{3}{2},n+\frac{7}{2};\rho)}
{_2F_1(n,\frac{1}{2},n+\frac{5}{2};\rho)}
\end{eqnarray}
with $\xi=\frac{Q_0^2}{4m_c^2}$ and $\rho=\frac{\xi}{1+\xi}$.  The factors multiplying the
condensate in $\phi^4_b$ and $\phi^4_c$ in eq.(\ref{phibc}) are
defined such that at $n \rightarrow \infty~$
$c_n\sim b_n$.   Moreover, at this limit, one notes that the
contribution from dimension 4  is proportional to
$\phi_b^4+\phi_c^4$, which in our kinematical limit originates
from the following operator form.
\begin{eqnarray}
\phi_b^4+\phi_c^4= \frac{16 \pi^2}{9}
{ \langle \frac{\alpha}{\pi} G^a_{k0}G^a_{k0} \rangle \over
 (4m_c^2)^2 },
\end{eqnarray}
where the operator does not have the trace part of the $00$ index.
That is, the leading mass shift is proportional to the color
electric field squared $E^2$ and the effect coming from the
magnetic field squared $B^2$ disappears.  This is
consistent with the non-relativistic picture at the infinite
quark mass limit\cite{Luk92}, where the Zeeman effect is
higher order in $\alpha_s$ compared to the Stark effect.

The  dimension 6 operators contribute to the moment as follows,
\begin{eqnarray}
\Delta M^6_n(\xi) & =& A_n^V(\xi)\Big[\, s_n(\xi) \phi^6_{s}+t_n(\xi) \phi^6_{t}
+x_n(\xi) \phi^6_{x} \nonumber \\
&& +y_n(\xi) \phi^6_{y}+z_n(\xi) \phi^6_{z}+g_{4n}(\xi) \phi^6_{g_4} \Big],
\end{eqnarray}
where
\begin{eqnarray}
\phi^6_s&=&\frac{4\pi^2}{3\cdot 1080}
\frac{\langle \frac{\alpha_s}{\pi} G^a_{\mu\nu} G^a_{\mu\nu;\kappa\kappa} \rangle}{(4m_c^2)^3}
\nonumber\\
\phi^6_{t}&=&\left(\frac{2}{3}\right)\frac{4\pi^2}{3\cdot 1080}
\frac{\langle \frac{\alpha_s}{\pi} G^a_{\alpha\kappa} G^a_{\alpha\lambda;\lambda\kappa}
\rangle}{(4m_c^2)^3},
\label{eqn:6s}
\end{eqnarray}
come from the scalar operators
and
\begin{eqnarray}
\phi^6_{x}&=&\left( \frac{9m_N^2}{2}\right)\frac{4\pi^2}{3\cdot 1080}
\frac{X}{(4m_c^2)^3} \frac{\rho}{2m_N}
\nonumber\\
\phi^6_{y}&=&\left( \frac{3m_N^2}{2}\right)\frac{4\pi^2}{3\cdot 1080}
\frac{Y}{(4m_c^2)^3} \frac{\rho}{2m_N}
\nonumber\\
\phi^6_{z}&=&\left( \frac{3m_N^2}{2}\right)\frac{4\pi^2}{3\cdot 1080}
\frac{Z}{(4m_c^2)^3} \frac{\rho}{2m_N}
\nonumber\\
\phi^6_{g_4}&=&( 10m_N^4)\frac{4\pi^2}{3\cdot 1080}
\frac{G_4}{(4m_c^2)^3} \frac{\rho}{2m_N},
\label{eqn:6t}
\end{eqnarray}
are from tensor operators.
Note here again that we have put in the prefactors in eq.(\ref{eqn:6s}) and eq.(\ref{eqn:6t})
such that in the large $n$ limit, all the Wilson coefficients
become the same.
Specifically, the Wilson coefficients are,
\begin{eqnarray}
s_n(\xi)&=&\sigma_n(\xi) f_n(0,0;120,-310,258,-72)\nonumber\\
t_n(\xi)&=&s_n(\xi)+\sigma_n(\xi)f_n(-120,-60;-300,1320,-720)\nonumber\\
x_n(\xi)&=&s_n(\xi)+\sigma_n(\xi)f_n(-70,25;245,-80,-6)\nonumber\\
y_n(\xi)&=&s_n(\xi)+\sigma_n(\xi)f_n(270,-45;-1305,1680,-666)\nonumber\\
z_n(\xi)&=&s_n(\xi)+\sigma_n(\xi)f_n(390,-225;-1245,880,-306)\nonumber\\
g_{4n}(\xi)&=&s_n(\xi)+\sigma_n(\xi)f_n(-210,-33;-21,48,-18),
\end{eqnarray}
where
\begin{eqnarray}
&&\sigma_n(\xi)=\left( \frac{-8}{1+\xi} \right) \frac{n(n+2)}{(2n+5)(2n+7)}
\end{eqnarray}
and
\begin{eqnarray}
&&f_n(c_1,c_2;a_2,a_3,\cdots,a_k)\nonumber\\
&\equiv &
\left( \begin{array}{c} c_1,c_2,a_2,a_3,\cdots,a_k\end{array} \right)
\left( \begin{array}{cc}
(n+3)\frac{_2F_1(n+1,\frac{1}{2},n+\frac{9}{2}; \rho)}{_2F_1(n,\frac{1}{2},n+\frac{5}{2}; \rho)}
\\
-2(2n+7)\frac{_2F_1(n+1,\frac{1}{2},n+\frac{7}{2}; \rho)}{_2F_1(n,\frac{1}{2},n+\frac{5}{2}; \rho)}
\\
(n+3)(n+4)\frac{_2F_1(n+1,-\frac{1}{2},n+\frac{9}{2}; \rho)}{_2F_1(n,\frac{1}{2},n+\frac{5}{2}; \rho)}
\\
\frac{(n+3)(n+4)(n+5)}{2!}
\frac{_2F_1(n+1,-\frac{3}{2},n+\frac{9}{2}; \rho)}{_2F_1(n,\frac{1}{2},n+\frac{5}{2}; \rho)}
\\
\vdots
\\
\frac{(n+3)(n+4)\cdots(n+k+2)}{(k-1)!}
\frac{_2F_1(n+1,-k+\frac{3}{2},n+\frac{9}{2}; \rho)}{_2F_1(n,\frac{1}{2},n+\frac{5}{2}; \rho)}
\end{array}\right).
\end{eqnarray}
These functions can be read off from the polarizations which are
expressed in terms of linear combinations of the $J_N(y)$'s.
See  appendix \ref{hyp}.

\subsection{estimation of matrix elements}

Here, we summarize the parameters and matrix elements
appearing in our sum rule.
As in ref.\cite{Kli99,Rei81}, we will choose the normalization
point to be $Q^2=4m_c^2$; i.e. $\xi=1$.  Hence, we will use\cite{Kli99}
\begin{eqnarray}
\label{alphas}
\alpha_s(8m_c^2) =   0.21,~~
m_c   = 1.24\,{\rm GeV}.
\end{eqnarray}
 For the matrix elements, we will use linear density approximation
in eq.(\ref{linear}) with $m_N=0.93$ GeV and the nuclear matter density to be
$\rho_0=0.17/{\rm fm}^3$.

\subsubsection{scalar operators}

There are  one scalar operator in dimension 4 and two in dimension 6.

\begin{enumerate}

\item
$\langle \frac{\alpha_s}{\pi} G^2 \rangle_\rho$

 The density dependence of the scalar gluon condensate is obtained from
the trace anomaly relation in the chiral limit which to leading order
in $\alpha_s$ is, $\theta_\mu^\mu=-\frac{9\alpha_s}{8 \pi}
G_{\mu \nu}^a G_{\mu \nu}^a$.    Taking the nucleon expectation value, we find,
\begin{eqnarray}
\label{scalar1}
\langle \frac{\alpha_s}{\pi} G^2 \rangle_\rho&=&
 \langle \frac{\alpha_s}{\pi} G^2 \rangle_0 -\frac{8}{9}
m_N^0 \rho\nonumber\\
&\simeq&(0.35\, {\rm GeV})^4
\left(1 -a_1 \frac{\rho}{\rho_0} \right).
\end{eqnarray}
We used $m_N^0=750$ MeV for the
nucleon mass in the chiral limit\cite{Bor96} to estimate
\begin{eqnarray}
a_1=\frac{8m_N^0 \rho_0}{9\langle \frac{\alpha_s}{\pi} G^2 \rangle_0}\simeq 0.058.
\end{eqnarray}

\item $\langle \frac{\alpha_s}{\pi} G^a_{\mu\nu} G^a_{\mu\alpha;\alpha\nu}
\rangle_\rho=(-\frac{1}{4\pi^2})\langle g^4 j^2 \rangle_\rho$

Where we have used the equation of motion in eq.(\ref{em}) to rewrite the gluon
operator in terms of quark operators.

This is a four quark operator. Hence to estimate the nuclear matter
expectation value,
we use ground state saturation hypothesis, where one can factor
out the four quark operator in terms of independent two quark operator and
take their ground state expectation value\cite{Shi79,Jin93}.  This is a
generalization of   vacuum dominance hypothesis(VDH) in the vacuum\cite{Shi79}.
Hence,

\begin{eqnarray}
\label{scalar2}
\langle j^2 \rangle_\rho
&=&-\frac{N_c^2 -1}{2 \cdot 4^2 \cdot N_c} \Big\{
\langle {\bar q}q \rangle_\rho^2 {\rm Tr}[\gamma_\mu \gamma_\mu]\nonumber\\
&&+\langle
{\bar q}q \rangle_\rho
\langle {\bar q}\gamma_\kappa q \rangle_\rho {\rm Tr}[\gamma_\mu \gamma^\kappa
\gamma_\mu+
\gamma^\kappa \gamma_\mu \gamma_\mu]\nonumber\\
&&+\langle {\bar q}\gamma_\kappa q \rangle_\rho \langle {\bar q}\gamma_\lambda q \rangle_\rho
{\rm Tr}[\gamma^\kappa \gamma_\mu \gamma^\lambda \gamma_\mu]\Big\}
\nonumber\\
&=&(-\frac{1}{3})\left[
g_{\mu\mu}\langle {\bar q}q \rangle_\rho^2 +2
\langle {\bar q}\gamma_\mu q \rangle_\rho \langle {\bar q}\gamma_\mu q \rangle_\rho-g_{\mu\mu}
\langle {\bar q}\gamma_\lambda q \rangle_\rho \langle {\bar q}\gamma_\lambda q \rangle_\rho
\right]\nonumber\\
&=&(-\frac{1}{3})\left[
4\langle {\bar q}q \rangle_\rho^2-2
\langle {\bar q}\gamma_\lambda q \rangle_\rho \langle {\bar q}\gamma_\lambda q \rangle_\rho
\right] \nonumber \\
&\simeq& -\frac{4}{3}\cdot \left\{ \langle {\bar q}q \rangle_0^2 +
\frac{2\sigma_N \langle {\bar q}q  \rangle_0}{m_u +m_d} \rho \right\}
\nonumber \\
& =&-\frac{4}{3}\langle {\bar q}q \rangle_0^2
\left( 1 - a_2\frac{\rho}{\rho_0} \right)\simeq
-(0.24\, {\rm GeV})^6\left( 1 - a_2\frac{\rho}{\rho_0} \right),
\end{eqnarray}
with
\begin{eqnarray}
a_2=\frac{-2\sigma_N \rho_0}{(m_u+m_d)\langle {\bar q}q \rangle_0}\simeq 0.69.
\end{eqnarray}

In getting this, we first note,
\begin{eqnarray}
\label{dim3}
\langle {\bar q}\gamma_\mu q \rangle_\rho=\langle {\bar q}\not\!u q \rangle_\rho u_\mu=
\langle {q}^\dagger q \rangle_\rho u_\mu \propto \rho \,g_{\mu 0},
\end{eqnarray}
where we have introduced the four vector $u_\mu$ to represent
the nuclear matter, which in our case is $u=(1,0,0,0).$
We used this in eq.(\ref{scalar2}) and neglected
terms proportional to $\rho^2$ .
Other values  taken in eq.(\ref{scalar2})
are, $\langle {\bar q}q \rangle_0=(-0.23~{\rm GeV})^3$ and
 $ \frac{2\sigma_N}{m_u+m_d}=0.09/0.014$\cite{HL92}.
Finally, we have,
\begin{eqnarray}
\langle \frac{\alpha_s}{\pi} G^a_{\mu\nu} G^a_{\mu\alpha;\alpha\nu} \rangle_\rho
&=&\left(-\frac{1}{4\pi^2}\right)\langle g^4 j^2 \rangle_\rho\nonumber\\
&=&\left(-\frac{1}{4\pi^2}\right)\langle g^4 j^2 \rangle_0
\left( 1 - a_2 \frac{\rho}{\rho_0} \right)\nonumber\\
&=&(0.23\, {\rm GeV})^6
\left( 1 - a_2 \frac{\rho}{\rho_0} \right),
\end{eqnarray}

It should be noted that here we have used a larger value of $\alpha_s=0.7$
compared to the perturbative value in eq.(\ref{alphas})\cite{Nik83}.

\item
$\langle \frac{\alpha_s}{\pi} G^a_{\mu\nu} G^a_{\mu\nu;\alpha\alpha} \rangle_\rho$

Using the identities in the appendix, one can rewrite the operator in terms of
the three gluon operator and the four quark operator.
 $\frac{1}{2\pi^2}\langle g^3G^3 -g^4j^2 \rangle_\rho$.  For the four quark
operator, we use the previous result.  For the density dependence of the
three gluon operator, we assume that the following ratio is constant to
linear order in density.
\begin{eqnarray}
R=\frac{\langle g^3G^3 \rangle_0^{2/3}}{\langle g^2G^2 \rangle_0}=
\frac{\langle g^3G^3 \rangle_\rho^{2/3}}{\langle g^2G^2 \rangle_\rho},
\end{eqnarray}
This, together with the previous results on the density dependence of the
gluon condensate and the four quark operator, we find,
\begin{eqnarray}
\langle \frac{\alpha_s}{\pi} G^a_{\mu\nu} G^a_{\mu\nu;\alpha\alpha} \rangle_\rho
&=&\frac{1}{2\pi^2} \langle g^3G^3 -g^4 j^2 \rangle_0
\nonumber\\
&=&\frac{1}{2\pi^2}\langle g^3G^3 \rangle_0 ( 1-\frac{3}{2}a_1 \frac{\rho}{\rho_0})
-\frac{1}{2\pi^2}\langle g^4j^2 \rangle_0 ( 1-a_2 \frac{\rho}{\rho_0} )\nonumber\\
&=&(0.60\, {\rm GeV})^6
\left( 1-a_3 \frac{\rho}{\rho_0} \right)
\end{eqnarray}
where
\begin{eqnarray}
a_3 =
\frac{\frac{3}{2}a_1 \langle g^3G^3 \rangle_0 -a_2 \langle g^4 j^2 \rangle_0}
{\langle g^3G^3 -g^4 j^2 \rangle_0}
\simeq 0.154
\end{eqnarray}
\end{enumerate}

\subsubsection{twist-2 operators}
 For the twist-2 gluonic operators, we have
\begin{eqnarray}
G_{2n}=-(-i)^{2n-2}\frac{\alpha_s}{\pi}A_{2n}^G,
\end{eqnarray}
where
\begin{eqnarray}
A^G_n(\mu^2)=2 \int^1_0 dx\,x^{n-1}G(x,\mu^2).
\end{eqnarray}
We take $A^G_2(8m_c^2)\simeq 0.9$ and $A^G_4(8m_c^2)\simeq 0.02$~\cite{Glu92}.

\subsubsection{twist-4 operators: dimension 6}

Here, there are three independent operators given in eq.(\ref{opall})
and contributing to
$\phi^6_x,\phi^6_y,\phi^6_z$.

\begin{enumerate}

\item
$\langle N|\frac{\alpha_s}{\pi} G^a_{\kappa\lambda}
G^a_{\kappa\lambda;\mu\nu} |N\rangle$

This operator can be considered to be the second moment of the gluon
condensate.  For this operator we assume
\begin{eqnarray}
\langle N|\frac{\alpha_s}{\pi} G^a_{\kappa\lambda} G^a_{\kappa\lambda;\mu\nu} |N\rangle &=&
(p_\mu p_\nu -\frac{1}{4} g_{\mu\nu} m_N^2 )\times (-)\frac{A^G_4}{A^G_2}
\langle N|\frac{\alpha_s}{\pi}G^2 |N\rangle
\nonumber\\
&=&(p_\mu p_\nu -\frac{1}{4} g_{\mu\nu} m_N^2 )\times \frac{0.02}{0.9}\cdot \left(
+2m_N\frac{8}{9} m^0_N  \right)
\nonumber\\
&=&(p_\mu p_\nu -\frac{1}{4} g_{\mu\nu} m_N^2  )\times (\frac{0.32}{8.1}) m_N m_N^0.
\end{eqnarray}
Therefore,
\begin{eqnarray}
X=(\frac{0.32}{8.1}) m_N m_N^0.
\end{eqnarray}

\item
$\langle N| G_{\mu \kappa} G_{\nu \lambda;\lambda\kappa}|N\rangle
= (-\frac{1}{4\pi^2})\langle g^4 j_\mu j_\nu \rangle_\rho$

Here, we have used the equation motion again to change this operator
to four quark operator.  However, this operator is traceless and
symmetric.  Consider the nuclear matter expectation value of this
operator,  it becomes,
${\cal S.T.}\langle g^4 j^a_\mu j^a_\nu \rangle_\rho=\langle g^4 j^a_\mu j^a_\nu \rangle_\rho -
\frac{1}{4}g_{\mu\nu} \langle g^4 j^a_\lambda j^a_\lambda \rangle_\rho$.
Assume the ground state saturation hypothesis again,
\begin{eqnarray}
\langle j^a_\mu j^a_\nu \rangle_\rho
&=&-\frac{N_c^2 -1}{2\cdot 4^2 \cdot N_c} \Big\{
\langle {\bar q}q \rangle_\rho^2 {\rm Tr}[\gamma_\mu \gamma_\nu]\nonumber\\
&&
+\langle {\bar q}q \rangle_\rho
\langle {\bar q}\gamma_\kappa q \rangle_\rho {\rm Tr}[\gamma_\mu \gamma^\kappa \gamma_\nu+
\gamma^\kappa \gamma_\mu \gamma_\nu]\nonumber\\
&&
+\langle {\bar q}\gamma_\kappa q \rangle_\rho \langle {\bar q}\gamma_\lambda q \rangle_\rho
{\rm Tr}[\gamma^\kappa \gamma_\mu \gamma^\lambda \gamma_\nu]\Big\} \nonumber\\
&=&(-\frac{1}{3})\left[
g_{\mu\nu}\langle {\bar q}q \rangle_\rho^2 +2
\langle {\bar q}\gamma_\mu q \rangle_\rho \langle {\bar q}\gamma_\nu q \rangle_\rho-g_{\mu\nu}
\langle {\bar q}\gamma_\lambda q \rangle_\rho \langle {\bar q}\gamma_\lambda q \rangle_\rho
\right]\nonumber\\
\langle j^a_\nu j^a_\nu \rangle_\rho
&=&(-\frac{1}{3})\left[
4\langle {\bar q}q \rangle_\rho^2-2
\langle {\bar q}\gamma_\lambda q \rangle_\rho \langle {\bar q}\gamma_\lambda q \rangle_\rho
\right].
\end{eqnarray}

So
\begin{eqnarray}
{\cal S.T.}\langle j^a_\mu j^a_\nu \rangle_\rho
&=&-\frac{2}{3}\langle {\bar q}\gamma_\mu q \rangle_\rho
\langle {\bar q}\gamma_\nu q \rangle_\rho+
\frac{1}{6}g_{\mu\nu}
\langle {\bar q}\gamma_\lambda q \rangle_\rho \langle {\bar q}\gamma_\lambda q \rangle_\rho.
\end{eqnarray}
Using eq.(\ref{dim3}),
\begin{eqnarray}
{\cal S.T.}\langle j^a_\mu j^a_\nu \rangle_\rho&=&
\frac{1}{6} \langle {q}^\dagger q \rangle_\rho^2 (u^2g_{\mu\nu}-4u_\mu u_\nu)
\sim \rho^2,
\end{eqnarray}
Hence,
\begin{eqnarray}
Y=0
\end{eqnarray}

\item
$\langle N|\frac{\alpha_s}{\pi} G^a_{\mu\kappa} G^a_{\kappa\lambda;\lambda\nu}|N \rangle$

Here we use the equation of motion once.  Then one finds that the
operator has similar structure to a twist-4 quark gluon mixed operator
appearing in the twist-4 contribution in deep inelastic
scattering\cite{Jaffe}.
\begin{eqnarray}
\langle N| G^a_{\mu\kappa} G^a_{\kappa\lambda;\lambda\nu} |N\rangle
&=&\langle N|
{\bar q}\gamma_\kappa [D_\nu,\,G_{\mu\kappa}]q |N\rangle\nonumber\\
&\simeq&
\langle N|i{\bar q}\{ D_\mu,\,^*F_{\nu\lambda}\}\gamma^\lambda \gamma_5 q |N\rangle
\nonumber\\
&=&\frac{3}{2}\langle N|i{\bar u} \gamma_\kappa \gamma_5
\{ D_\nu,\,^*F_{\mu\kappa}\} u |N\rangle\nonumber\\
&=&(p_\mu p_\nu-\frac{1}{4}g_{\mu\nu} m_N^2)\cdot \frac{3}{2}K^g_u.
\end{eqnarray}
Here, $K^g_u$ is a parameter introduced in ref.\cite{Cho93,Lee94} to fit the
available deep inelastic scattering data.  The fit gives,
$-0.3\,{\rm GeV}^2<K^g_u<-0.2\,{\rm GeV}^2$.  With this one finds,
\begin{eqnarray}
Z \simeq (0.097 \,{\rm GeV})^2.
\end{eqnarray}
\end{enumerate}

In Table \ref{tab:conden}, we summarize values of the re-scaled
condensates  in Eq. (\ref{eqn:4}),(\ref{eqn:6s}) and (\ref{eqn:6t})
in the vacuum and at normal nuclear matter density.

\begin{table}
\caption{Expectation values of normalized gluon operators
in the vacuum and at normal nuclear matter density\label{tab:conden}}
\begin{tabular}{|l||l|l|l|l|}
\hline
Element  & $\phi^4_b$ & $\phi^4_c$ & $\phi^6_s$ & $\phi^6_t$
\\
\hline
in vacuum & $1.67\times10^{-3}$ & $0$ & $1.66\times10^{-7}$ & $1.65\times10^{-8}$
\\
in matter & $1.58\times10^{-3}$ & $-1.53\times10^{-5}$ & $1.28\times10^{-7}$ & $6.50\times10^{-9}$
\\
{\em change} & $-9.70\times10^{-5}$ & $-1.53\times10^{-5}$ & $-3.88\times10^{-8}$ & $-1.00\times10^{-8}$
\\
\hline\hline
Element  & $\phi^6_x$ & $\phi^6_y$ & $\phi^6_z$ & $\phi^6_{g_4}$
\\
\hline
in vacuum & $0$ & $0$ & $0$ & $0$
\\
in matter & $1.67\times10^{-8}$ & $0$ & $4.27\times10^{-10}$ & $-1.69\times10^{-9}$
\\
{\em change} & $1.67\times10^{-8}$ & $0$ & $4.27\times10^{-10}$ & $-1.69\times10^{-9}$
\\
\hline
\end{tabular}
\end{table}

\subsection{numerical analysis}

\begin{figure}[htb]
\begin{minipage}[t]{77mm}
\vbox to 2.3in{\vss
   \hbox to 1.3in{\includegraphics{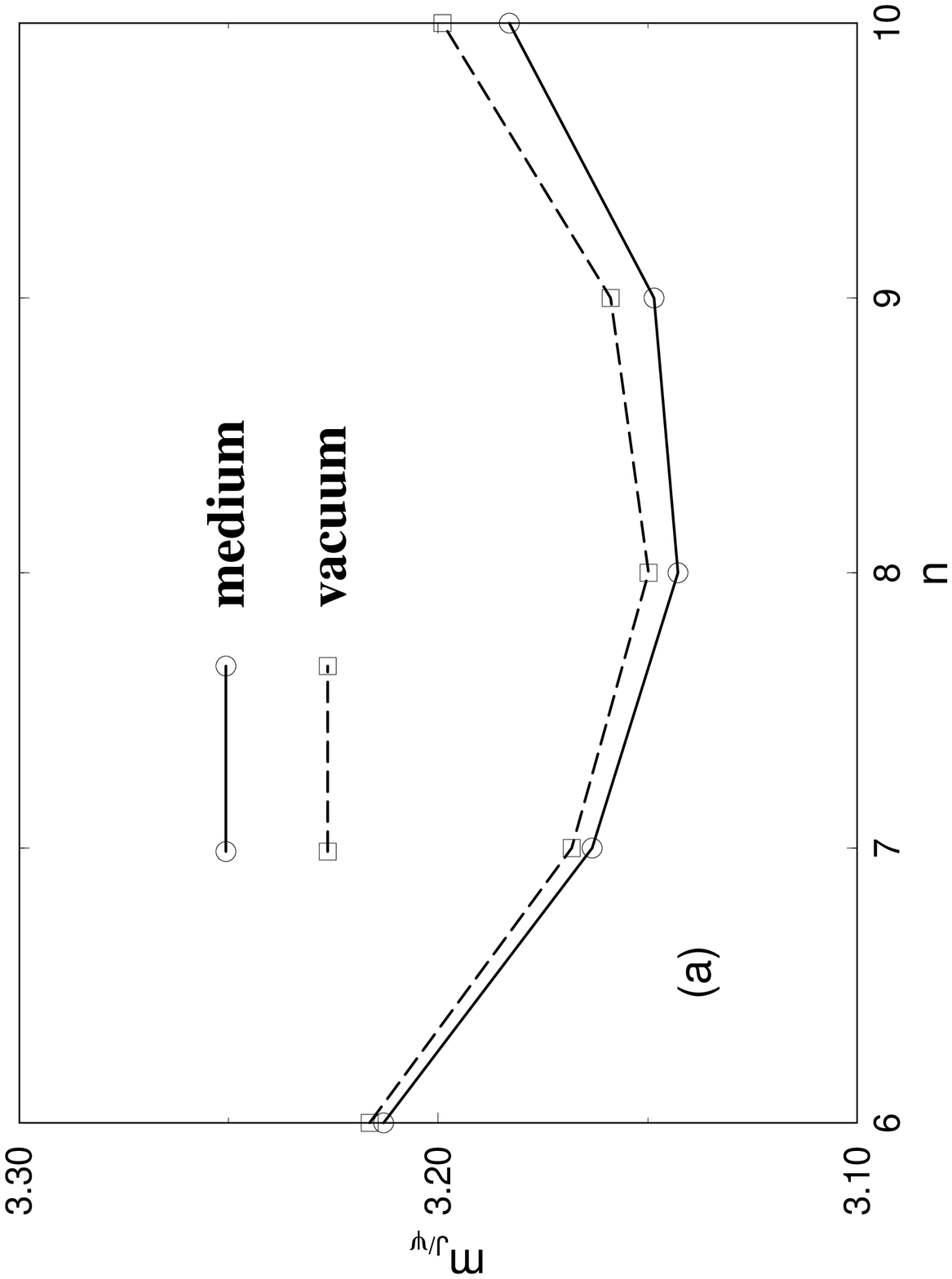}\hss}}
\end{minipage}
\hspace{\fill}
\begin{minipage}[t]{77mm}
\vbox to 2.3in{\vss
   \hbox to 1.3in{\includegraphics{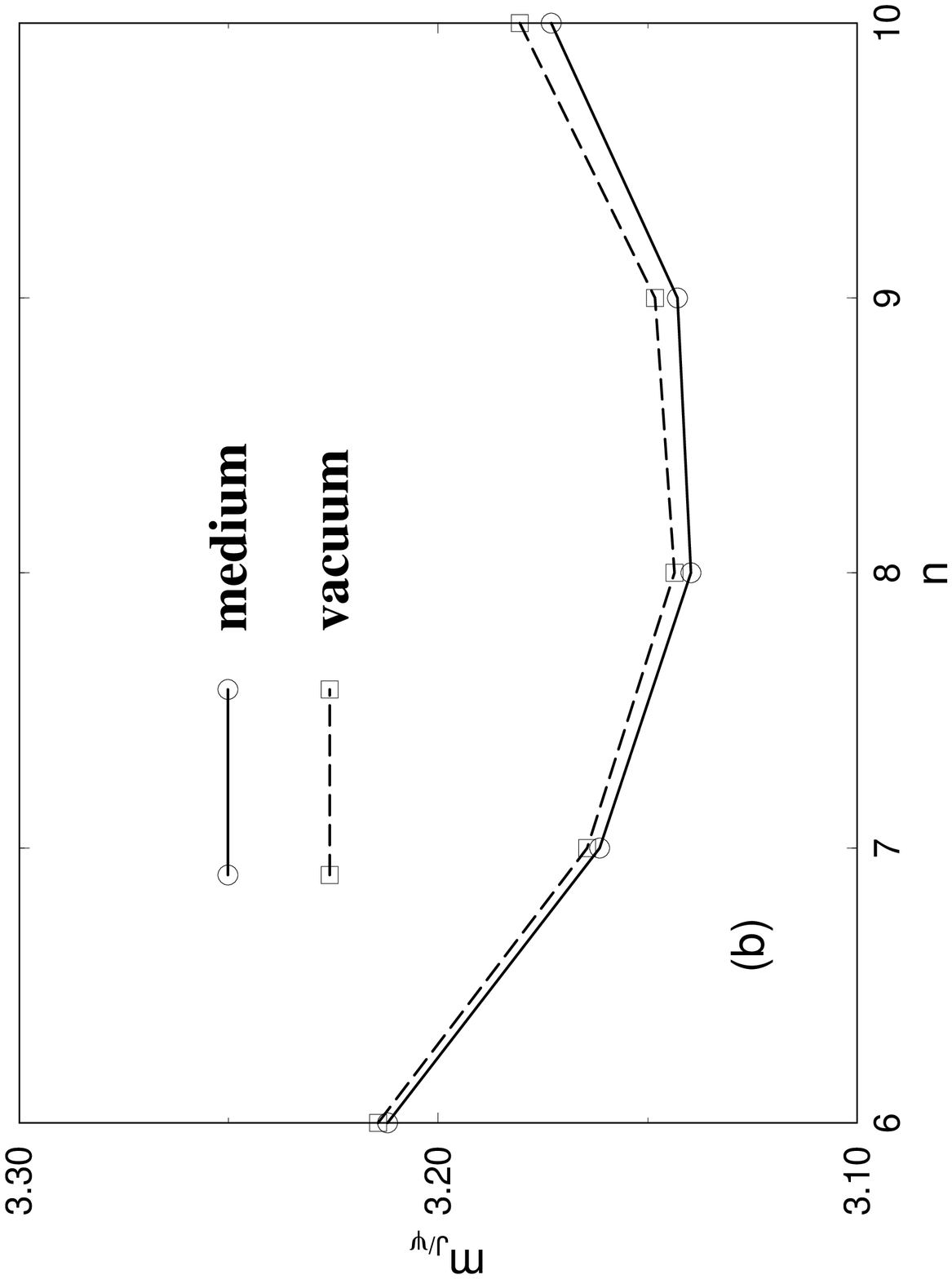}\hss}}
\end{minipage}
\caption{The mass of $J/\psi$ in GeV determined from the plot of eq.(\ref{massshift})
for different $n$ at $\xi=1$. We compare the result in normal
nuclear matter (solid line)  with the vacuum result (dashed line).
(a) refers to the previous calculation, which includes
only dimension 4 operators.  (b) is the present calculation
which includes all the dimension 6 operators.\label{fig:mass}}
\end{figure}

The polarization function $\Pi$ satisfies the following energy dispersion
relation.
\begin{eqnarray}
\Pi(\omega^2)=\frac{1}{\pi} \int ds { {\rm Im}\Pi(s) \over s-\omega^2 }.
\end{eqnarray}
In the vacuum, the spectral density (Im$\Pi(s)$) consists of the $J/\psi$ pole,
 the contribution from the excited states $\psi'$ and the $ D \bar{D}$ continuum.
The contribution from the $J/\psi$ and the low-lying resonances can be approximated
by  delta functions inside the dispersion integral.  This is so because the
 $J/\psi$ mass lies below the continuum threshold and is dominated by electromagnetic
 decays.   This is also true in nuclear matter for a $J/\psi$ at rest.
The inelastic channels opening due to the scattering of the $J/\psi+N \rightarrow
\Lambda_c(2.28)+ \bar{D}(1.87)$ is forbidden by kinematics.  All other processes are OZI rule
violating.  Hence, the delta function approximation for the $J/\psi$ is also
good in the nuclear medium.
\begin{eqnarray}\label{imphi}
{\rm Im}\Pi(s)= f \delta(s-m_{J/\psi}^2) + f' \delta(s-m_{\psi'}^2)+..
\end{eqnarray}
The second term just represents a generic excited state contribution.
Substituting eq.(\ref{imphi}) into the dispersion relation, taking the
moments and taking the ratio between the $(n-1)$th and the $n$th moment, one
finds,
\begin{eqnarray}\label{nnminus}
{M_{n-1}(Q^2) \over M_n(Q^2)}=(m_{J/\psi}^2+Q^2)\times
\left( {1+  \delta_{n-1} \over 1+ \delta_n } \right)
\end{eqnarray}
where
\begin{eqnarray}
\delta_n=\frac{f'}{f} \left({ m_{J/\psi}^2+Q^2 \over m_{\psi'}^2+Q^2 }
\right)^{n+1}.
\end{eqnarray}
Substituting the vacuum values for the excited states, one finds that
the ratio ${1+  \delta_{n-1} \over 1+ \delta_n }$ goes to 1 for
$n>5$.
Hence, the mass is determined from the relation in eq.(\ref{nnminus}) at
 $Q^2=4m_c^2$.\footnote{The determination at different value of
 $Q^2$ was found to be not significant\cite{Hay99}}.
\begin{eqnarray}
\label{massshift}
m_{J/\psi}^2={M_{n-1}(4m_c^2) \over M_n(4m_c^2)}-
4m_c^2
\end{eqnarray}
As before\cite{Kli99}, we analyze this equation as a function of $n$.
In Fig.\ref{fig:mass}(a) , we plot the previous result, which includes the
contribution only up to dimension 4\cite{Kli99}.  In Fig.\ref{fig:mass}(b), we
plot the present result which includes the total dimension 6 contribution.
As can be seen from the comparison, the minimum occurs again at similar
 $n$ value and the change from the vacuum results are also similar.
We avoid fine tuning of the bare charm quark mass $m_c$ to fit the vacuum $J/\psi$
mass to its vacuum value, because we are only interested in the shift of the $J/\psi$ mass,
which is almost independent of this fine tuning.

Comparing the two graphs, one notes that the minimum occurs at the
same $n$ value and the graphs looks similar.
Comparing the changes from the vacuum curve and the medium curve at the
minimum point, we find
\begin{eqnarray}
\Delta m_{J/\psi} \simeq -4\, {\rm MeV}.
\end{eqnarray}
This mass shift is smaller than our previous calculation including
dimension 4 only.
The main reason for a smaller mass shift compared to including
dimension 4 only is as follows.  In the vacuum, the dimension 6 operators tend to
cancel the dimension 4 operators\cite{Nik83}.
This tendency is not only true but  more effective in the medium.
Therefore, including dimension 6 effects in
medium would effectively correspond to a smaller change
in the dimension 4 operators in our previous analysis in ref\cite{Kli99}.
This implies a smaller mass shift.
Below, we will try to elaborate on this point and to explain
the interplay of  each contribution.

\begin{itemize}

\item In Fig.\ref{fig:dimensions}, we plot the contributions from
dimension 4 and dimension 6 operators to the moments both in the vacuum and
in nuclear medium.  As expected, the dimension 6 contributions are
smaller compared to dimension 4\cite{Kli99}.  Nevertheless, one notes
that both in the vacuum and in the nuclear medium,
their contributions are opposite to each other so that the
contribution from dimension 6 operators tend to cancel the
contributions from dimension 4 operators.

\begin{figure}[htb]
\begin{minipage}[t]{77mm}
\vbox to 2.3in{\vss
   \hbox to 1.3in{\includegraphics{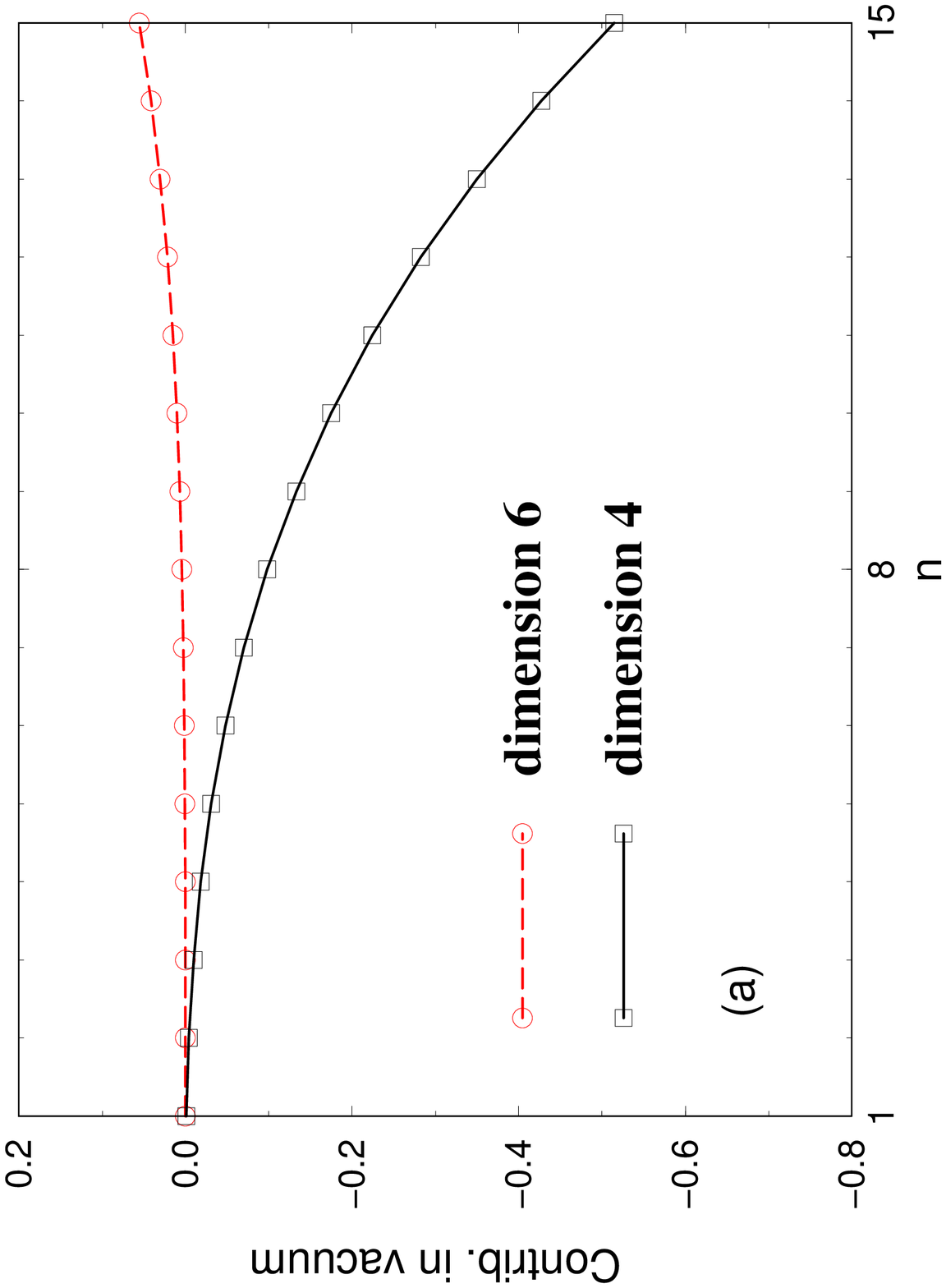}\hss}}
\end{minipage}
\hspace{\fill}
\begin{minipage}[t]{77mm}
\vbox to 2.3in{\vss
   \hbox to 1.3in{\includegraphics{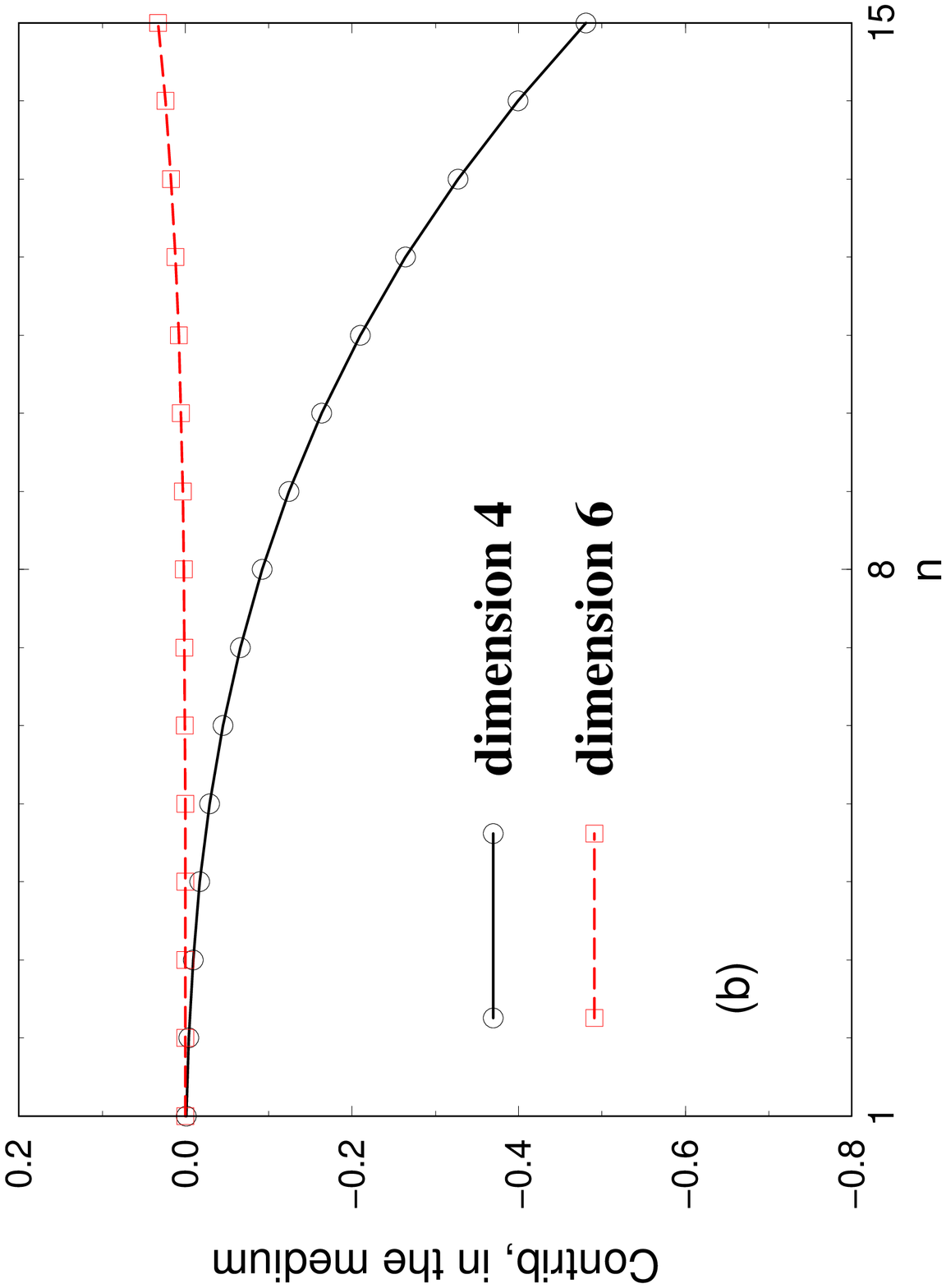}\hss}}
\end{minipage}
\caption{Contributions from dimension 4 and dimension 6 operators
to the moments both in the vacuum and in the nuclear medium in GeV$^{-2n}$.
\label{fig:dimensions}}
\end{figure}

\item The mass shift is coming from the changes in the
contributions from each dimension.  Hence, for comparison,
in Fig.\ref{fig:shifts},
we plot the contributions coming from each dimension in the
vacuum and in the nuclear medium.
One notes that effectively, each contributions becomes smaller in
nuclear medium.
Numerically, the ratio between the changes in dimension 4
operators and dimension 6 operators are roughly given by
(dimension 4 : dimension 6 $\simeq 3:1$).

\begin{figure}[htb]
\begin{minipage}[t]{77mm}
\vbox to 2.3in{\vss
   \hbox to 1.3in{\includegraphics{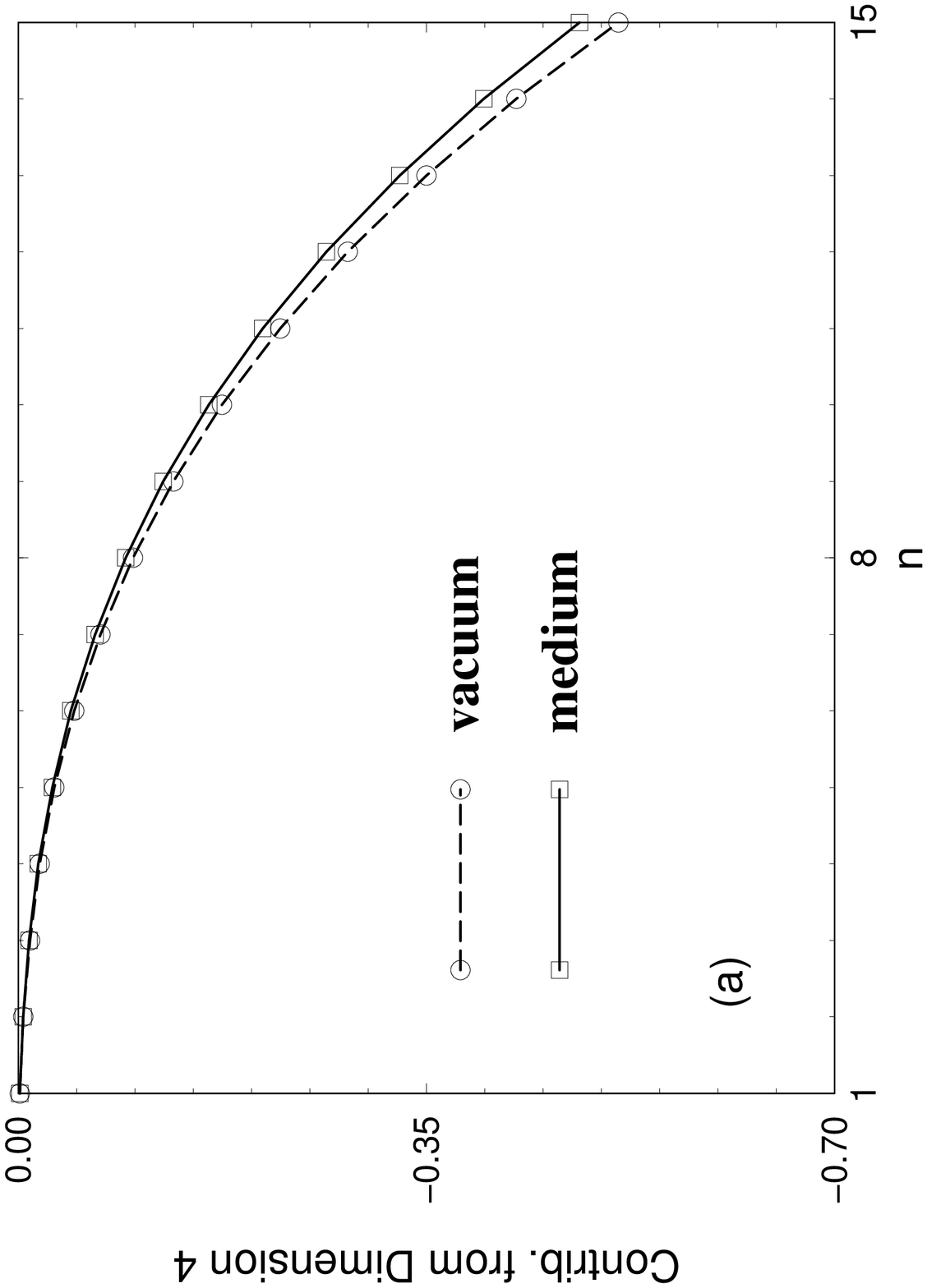}\hss}}
\end{minipage}
\hspace{\fill}
\begin{minipage}[t]{77mm}
\vbox to 2.3in{\vss
   \hbox to 1.3in{\includegraphics{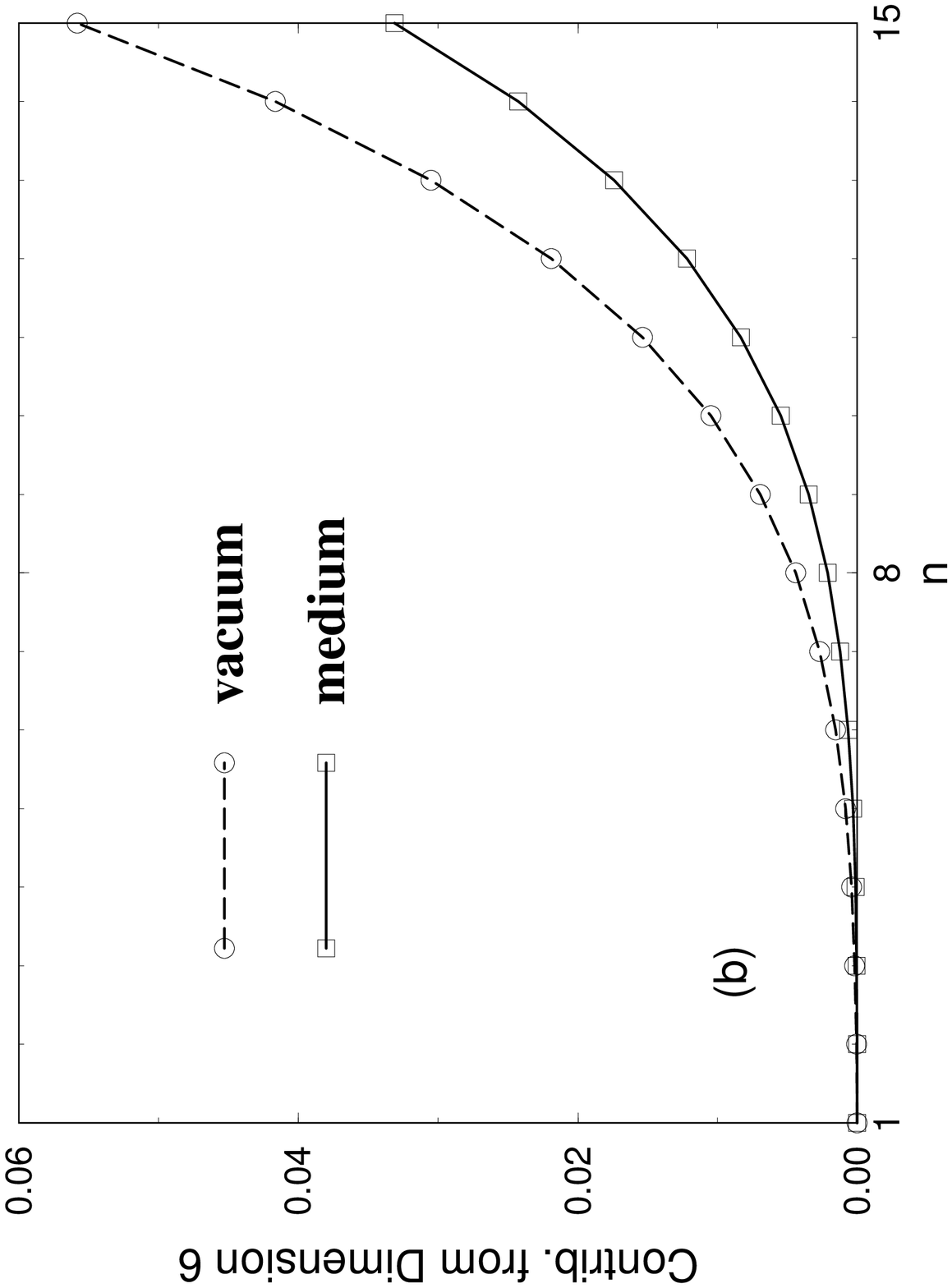}\hss}}
\end{minipage}
\caption{Contribution to the moments
in the vacuum and in the nuclear medium from (a) dimension 4 operators and
(b) dimension 6 operators.\label{fig:shifts}}
\end{figure}

\item The changes in dimension 4 operators are dominated by the changes in the
scalar gluon condensate\cite{Kli99}.  This is  evident  by comparing $\phi^4_b$ and
 $\phi^4_c$ in table 1.  The comparison between $\phi^4_b$ and $\phi^4_c$ is enough to compare
its contribution to the moments, because the Wilson coefficients multiplying
them are the same in the large $n$ limit.  Moreover,  $b_n=c_n\propto n^3$ and $b_n-c_n \propto
n$ such that $b_n \sim c_n$ at moderate $n$.

That is not quite so for the dimension 6 operators.  Here, the Wilson coefficients
multiplying $\phi^6_x,..$'s given in eq.(\ref{eqn:6t}) are also defined such that they are equal
in the large $n$ limit and are proportional to $n^5$.
However, the difference in the Wilson coefficients differ by only one power of $n$, that
is, the difference between Wilson coefficients are of order $n^4$.
Therefore, the Wilson coefficients are substantially different at the value of our interest,
 which is smaller than $n=10$, and become similar only at very large value of $n>100$.
  Hence, we will analyze each contribution from dimension 6 more in detail to identify
the important contributions.

In Fig.\ref{fig:dim6} (a), we have plotted the density dependent part of the scalar,
twist-2 and twist-4 contributions.   As can be seen, the important contributions are
the scalar operators.  And among the scalar operators, the contributions coming from
 $\phi^6_t$ is more important than that from $\phi^6_s$.  It should be noted that this
 is so because of the relatively large Wilson coefficients at small $n$
values.    The relative importance among the scalar operators
are the same also in the vacuum.
The next important set of operators are the twist-4 operators.   In Fig.\ref{fig:dim6} (b),
we plot the contributions from twist-4 operators, which has no vacuum expectation values.
The most important contribution is coming from $\langle G^a_{\mu \nu} G^a_{\mu \nu; \alpha
\beta}\rangle$, which contributes to $\phi^6_x$.  Here, the Wilson coefficients are similar
at least among the twist-4 operators and the relative importance can be
read off from
 Table \ref{tab:conden}.

\begin{figure}[htb]
\begin{minipage}[t]{77mm}
\vbox to 2.3in{\vss
   \hbox to 1.3in{\includegraphics{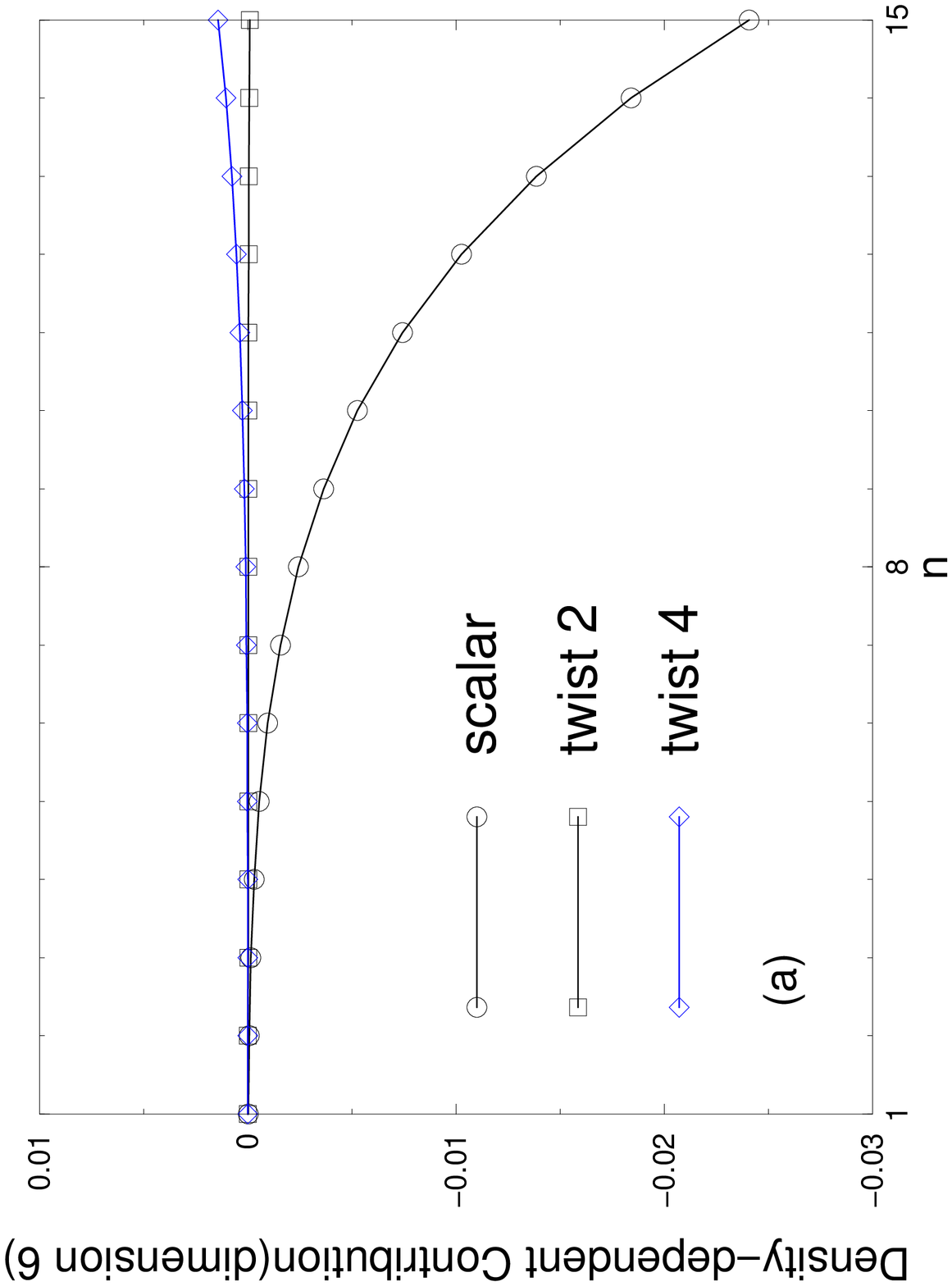}\hss}}
\end{minipage}
\hspace{\fill}
\begin{minipage}[t]{77mm}
\vbox to 2.3in{\vss
   \hbox to 1.3in{\includegraphics{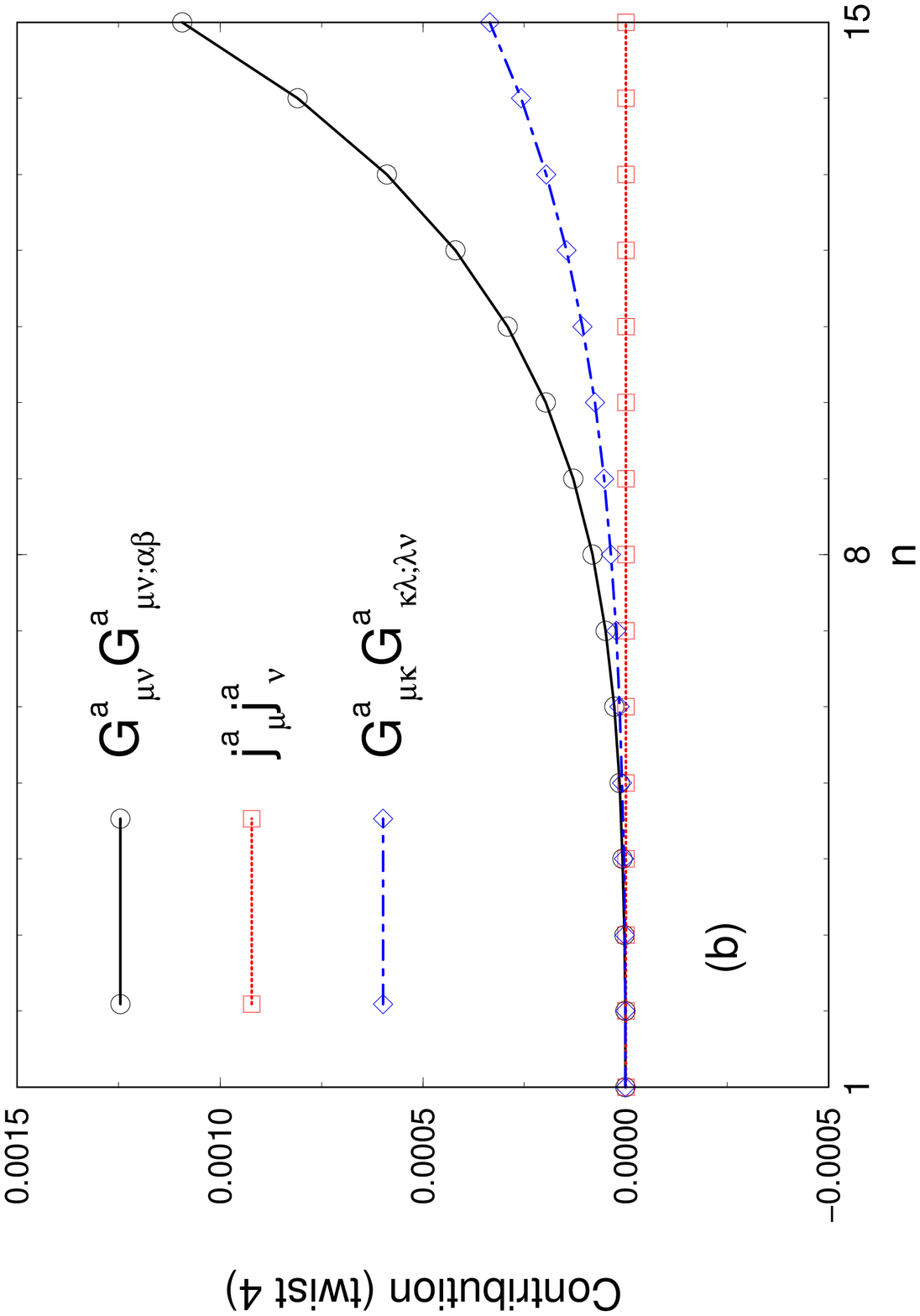}\hss}}
\end{minipage}
\caption{
(a) shows the contributions from the density
dependent part of the dimension 6 scalar , twist-4 and twist-2 operators to the moments.
(b) shows the individual contributions of the twist-4 operators.\label{fig:dim6}}
\end{figure}

\begin{figure}[htb]
\begin{minipage}[t]{77mm}
\vbox to 2.3in{\vss
   \hbox to 1.3in{\includegraphics{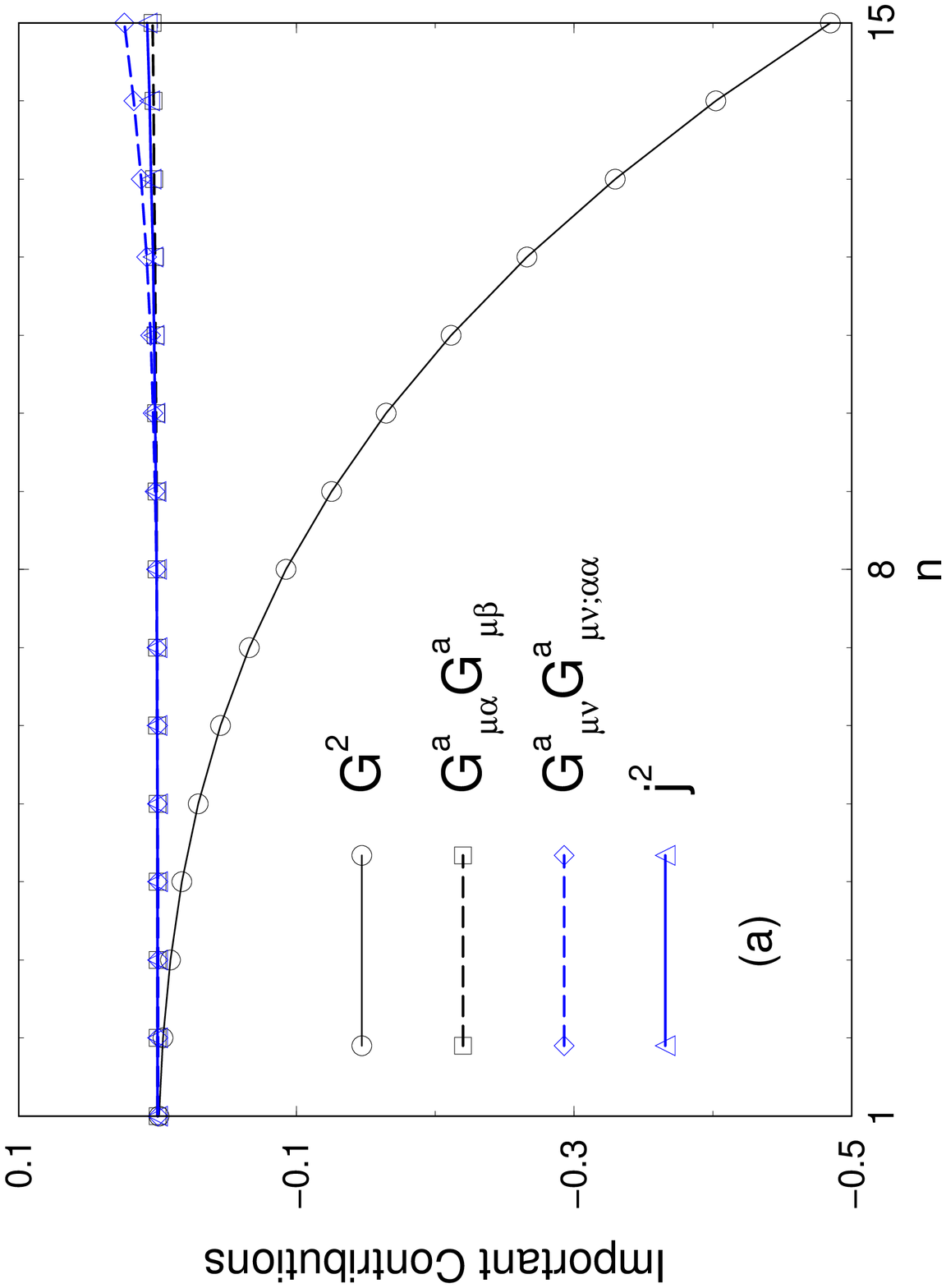}\hss}}
\end{minipage}
\hspace{\fill}
\begin{minipage}[t]{77mm}
\vbox to 2.3in{\vss
   \hbox to 1.3in{\includegraphics{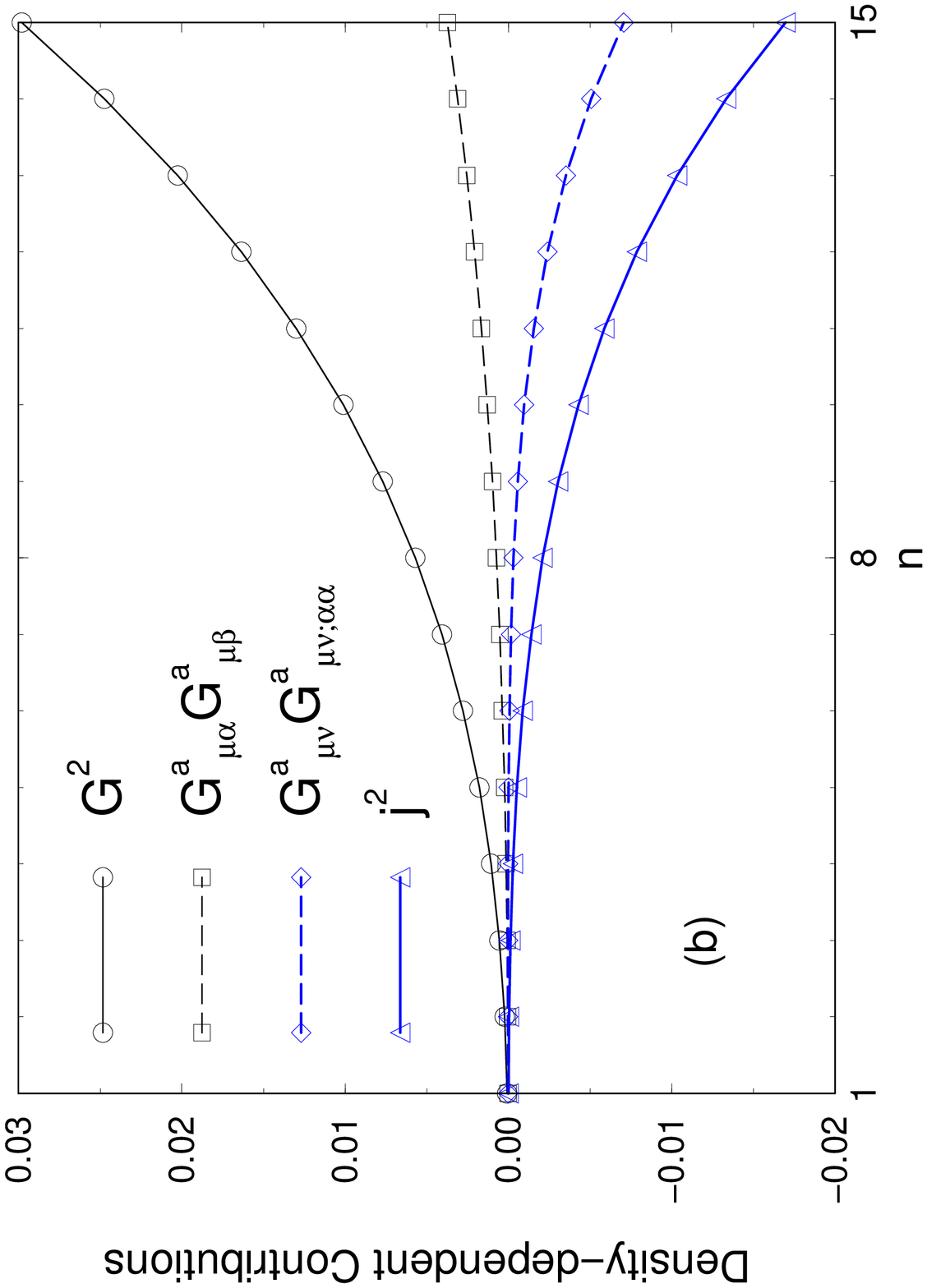}\hss}}
\end{minipage}
\caption{
(a) shows the important contributions to the moments coming from scalar operators
of dimension 4 and dimension 6.
(b) shows the density dependent part of (a), i.e. the
density dependent changes in the condensates. \label{fig:pures}}
\end{figure}

\item We finally compare the dominant contributions from dimension 4 and
dimension 6
operators to the moments.  In Fig. \ref{fig:pures} (a), we show the total
contributions
of the scalar dimension 4 and scalar dimension 6 operators to the moments.
As can be seen from the figure, the dominant contributions from  dimension 6
operators have opposite sign from dimension 4 operator and reduces its contribution.
This reduction is enhanced for the density dependent part.
This is shown in  Fig. \ref{fig:pures} (b).
The reason why the change in the dimension 6 operator are greater is clear.
The gluon condensate eq.(\ref{scalar1}), which is the dominant dim 4 operator, changes
in nuclear matter only by 6 \%, whereas, the scalar four quark condensate
eq.(\ref{scalar2}),
which is the dominant dim 6 operator, changes by almost 70\%.  Therefore, although
the Wilson coefficients are smaller for dimension 6 operator, the density dependent
part of dimension 6 operator are large such that it  cancels the
the contribution from dimension 4 operator non-trivially.

Hence, the reason why the mass shift gets smaller compared to just taking into
account dimension 4 operators is because the density dependent changes coming from
dimension 4 and dimension 6 operators  tend to cancel each other.   Therefore,
taking into account dimension 6 operators would effectively be equal to a smaller
change in the dimension 4 condensate in the previous result\cite{Kli99},
which would have given a smaller mass shift.

\end{itemize}

\section{Conclusion}

In this paper we have calculated the OPE of the correlation function between two
vector currents made of heavy quarks up to dimension 6 operators with any tensor
structure.
The formidable task was to categorize and calculate the corresponding Wilson coefficients of
dimension 6 twist-4 gluon operators.  This is a first attempt to establish the three
independent twist-4 gluon operators in terms of operator basis.

Using this result, we have applied our OPE to analyze the mass shift of $J/\psi$ in
nuclear medium using QCD moment sum rules for the heavy quark system.
 This is a
generalization of our previous result\cite{Kli99}, where we calculated the mass shift
using the OPE only up to dimension 4 operators.
Unfortunately, the nucleon expectation values of the dimension 6 operators are not
as reliable as the dimension 4 operators.
Nevertheless, using an order of magnitude
estimate for the matrix elements,
we find that the mass of $J/\psi$ would decrease by about $4$ MeV in the nuclear medium.
This is $3$ MeV smaller than the previous result on including only dimension 4 operators,
and shows that the dimension 6 effect is about  40\%  correction of the dimension 4
effects and goes in the opposite direction.   This result seems consistent with the
 notion
that the higher dimensional correction in the vacuum QCD sum rule for the heavy
quark system goes like $\left( G^2/m^4_c \right)^K$ in the
$r_n=\frac{M_n(\xi=0)}{M_{n-1}(\xi=0)}$ with alternating signs with
$K$\cite{Rad83}.
This also seems to be true in medium and the true mass shift is expected to lie
between $-4$ and $-7$ MeV.

The  resulting value of mass shift, is also  consistent with the more recent
estimates using a totally different approach\cite{Kaid92,Dim96}.
This is a result for a $J/\psi$ at
rest with respect to the nuclear medium.    However, since we have calculated the OPE
for a general external four momentum, our results can be easily and reliably
generalized to
study the moving $J/\psi$\cite{SL00} and also to finite temperature\cite{FHL90},
which would be also interesting in relation to the
ongoing discussion of $J/\psi$ suppression in RHIC due to a
comover model.

\section{Acknowledgement}

This work was supported by KOSEF grant number 1999-2-111-005-5, by the BK 21 project
of the Korean Ministry of Education and by the
Yonsei university research grant.
We would like to thank T. Hatsuda, A. Hayashigaki,
and W. Weise for useful discussions.  We particularly thank D. Kharzeev and
P. Morath for pointing out the difference in reference \cite{Luk92} and
\cite{Kaid92}.

\newpage\begin{appendix}
\section{Identities for spin-2 dimension 6 gluon operators\label{identity}}

The spin-2 dimension 6 gluon operators in eq.(\ref{gluonop}) are not independent.
The following relations holds among them.

\begin{eqnarray}
G^a_{\kappa\lambda;\mu} G^a_{\nu\kappa;\lambda}&=&G^a_{\kappa\lambda;\mu}
\cdot \frac{1}{2}(G^a_{\lambda\nu;\kappa} +G^a_{\nu\kappa;\lambda})
=-\frac{1}{2} G^a_{\kappa\lambda;\mu} G^a_{\kappa\lambda;\nu}\\
G^a_{\mu\kappa;\lambda} G^a_{\kappa\lambda;\nu}&=&
\frac{1}{2} \left( G^a_{\lambda\mu;\kappa} +G^a_{\mu\kappa;\lambda}\right)\cdot G^a_{\kappa\lambda;\nu}
=-\frac{1}{2} G^a_{\kappa\lambda;\mu} G^a_{\kappa\lambda;\nu}\\
G^a_{\mu\kappa;\nu} G^a_{\kappa\lambda;\lambda} &=&-G^a_{\mu\kappa;\nu\lambda} G^a_{\kappa\lambda}
=-(G^a_{\mu\kappa;\lambda\nu}+gf^{abc} G^b_{\mu\kappa} G^c_{\nu\lambda}) G^a_{\kappa\lambda}
\nonumber\\
&=&(-)g f G^3_{\mu\nu} +G^a_{\mu\kappa;\lambda} G^a_{\kappa\lambda;\nu}
\nonumber\\
&&{\rm Or,}\nonumber\\
g f G^3_{\mu\nu} &=&G^a_{\mu\kappa;\lambda} G^a_{\kappa\lambda;\nu}
-G^a_{\mu\kappa;\nu} G^a_{\kappa\lambda;\lambda}\nonumber\\
&=&-\left( \frac{1}{2} G^a_{\kappa\lambda;\mu} G^a_{\kappa\lambda;\nu}
+G^a_{\mu\kappa;\nu} G^a_{\kappa\lambda;\lambda} \right)
\\
G^a_{\kappa\lambda;\lambda}
G^a_{\mu\kappa;\nu}&=&-G^a_{\kappa\lambda;\lambda\nu}G^a_{\mu\kappa}=-\left(
G^a_{\kappa\lambda;\nu\lambda}+gf^{abc}
G^b_{\kappa\lambda} G^c_{\lambda\nu} \right)G^a_{\mu\kappa}
\nonumber\\
&=&+G^a_{\kappa\lambda;\nu} G^a_{\mu\kappa;\lambda} -g{\cal G}^3_{\mu\nu}
=G^a_{\mu\kappa;\nu} G^a_{\kappa\lambda;\lambda}\\
G^a_{\mu\kappa;\lambda} G^a_{\nu\lambda;\kappa}&=&-G^a_{\mu\kappa;\lambda\kappa}G^a_{\nu\lambda}
=-\left( G^a_{\mu\kappa;\kappa\lambda} +gf^{abc} G^b_{\mu\kappa} G^c_{\lambda\kappa}\right)G^a_{\nu\lambda}
\nonumber\\
&=&G^a_{\mu\kappa;\kappa} G^a_{\nu\lambda;\lambda} -g{\cal G}^3_{\mu\nu}
\nonumber\\
&=&G^a_{\mu\kappa;\kappa}G^a_{\nu\lambda;\lambda}
+\frac{1}{2} G^a_{\kappa\lambda;\mu}
G^a_{\kappa\lambda;\nu}+G^a_{\mu\kappa;\nu} G^a_{\kappa\lambda;\lambda}
\\
G^a_{\mu\kappa;\lambda} G^a_{\nu\kappa;\lambda}&=&-G^a_{\mu\kappa;\lambda\lambda} G^a_{\nu\kappa}
=-g\left( 2f^{abc} G^b_{\mu\lambda} G^c_{\kappa\lambda}
+j^a_{\mu;\kappa} -j^a_{\kappa;\mu} \right)G^a_{\nu\kappa}\nonumber\\
&=&(-2)g{\cal G}^3_{\mu\nu} +g^2j^a_\mu j^a_\nu
-g j^a_\kappa G^a_{\nu\kappa;\mu}\nonumber\\
&=&G^a_{\kappa\lambda;\mu} G^a_{\kappa\lambda;\nu}
+2G^a_{\mu\kappa;\nu}G^a_{\kappa\lambda;\lambda} +g^2 j^a_\mu
j^a_\nu-G^a_{\mu\kappa;\nu}G^a_{\kappa\lambda;\lambda}
\nonumber\\
&=&G^a_{\mu\kappa;\nu} G^a_{\kappa\lambda;\lambda}+g^2j^a_\mu j^a_\nu+
G^a_{\kappa\lambda;\mu} G^a_{\kappa\lambda;\nu}
\end{eqnarray}
Using these, one can reduce the operators in eq.(\ref{gluonop})
to the three independent operators in eq.(\ref{ope4}).

\section{Current Conservation in the Polarization Operators}

Let us consider the  polarization of vector currents defined
in eq.(\ref{polarization1}).  When the current is Hermitian,
i.e. $j_\mu =j_\mu ^\dag$,  the contribution from spin-2 dimension 4
operator can in general be written in terms of the following four independent
terms.
{\footnotesize\begin{eqnarray}
\Pi_{\mu\nu}(q)=\frac{1}{Q^2}\left[ \theta^1_{\mu\nu} +\frac{q^\alpha q^\beta q_\mu q_\nu}{Q^4}
\theta^2_{\alpha\beta}+\frac{q^\alpha q_\nu}{Q^2}\theta^3_{\mu\alpha}
+\frac{q^\alpha q_\mu}{Q^2}\theta^3_{\nu\alpha}+g_{\mu\nu}\frac{q^\alpha q^\beta}{Q^2}\theta^4_{\alpha\beta}
\right]
\end{eqnarray}}

Imposing current conservation, on finds,
\begin{eqnarray}
\theta^1_{\mu \nu} & = & \theta^3_{\mu \nu} \nonumber \\
\theta^2_{\mu \nu} & = & \theta^1_{\mu \nu}+\theta^4_{\mu \nu}
\end{eqnarray}

Similar relations holds for higher dimensional operator. This is the
operator form of showing the existence of two independent polarization directions in
medium for eq.(\ref{polarization1}).

Summarizing similar relations for different spins and dimensions, we have
\begin{enumerate}

\item dimension 4 and spin 2

{\scriptsize\begin{eqnarray}
\Pi^{4,2}_{\mu\nu}(q)=\frac{1}{Q^2}
\left[{\cal I}^2_{\mu\nu}+\frac{1}{Q^2} (q_\rho q_\mu {\cal I}^2_{\rho\nu}+
q_\rho q_\nu {\cal I}^2_{\rho\mu}) +
g_{\mu\nu} \frac{q_\rho q_\sigma}{Q^2} {\cal J}^2_{\rho\sigma}+
\frac{q_\mu q_\nu q_\rho q_\sigma}{Q^4}
({\cal I}^2_{\rho\sigma}+{\cal J}^2_{\rho\sigma})\right]
\nonumber\end{eqnarray}}

\item dimension 6 and spin 2

{\scriptsize\begin{eqnarray}
\Pi^{6,2}_{\mu\nu}(q)=\frac{1}{(Q^2)^2}
\left[I^2_{\mu\nu}+\frac{1}{Q^2} (q_\rho q_\mu I^2_{\rho\nu}+
q_\rho q_\nu I^2_{\rho\mu}) +
g_{\mu\nu} \frac{q_\rho q_\sigma}{Q^2} J^2_{\rho\sigma}+
\frac{q_\mu q_\nu q_\rho q_\sigma}{Q^4}
(I^2_{\rho\sigma}+J^2_{\rho\sigma})\right]
\nonumber\end{eqnarray}}

\item dimension 6 and spin 4

{\scriptsize\begin{eqnarray}
\Pi^{6,4}_{\mu\nu}(q)&=&
\frac{q_\kappa q_\lambda}{(Q^2)^3}
\bigg[~I^4_{\kappa\lambda\mu\nu}+\frac{1}{Q^2} (q_\rho q_\mu I^4_{\kappa\lambda\rho\nu}
+q_\rho q_\nu I^4_{\kappa\lambda\rho\mu})
+g_{\mu\nu} \frac{q_\rho q_\sigma}{Q^2} J^4_{\kappa\lambda\rho\sigma}\nonumber\\
&&
+\frac{q_\mu q_\nu q_\rho q_\sigma}{Q^4}
(I^4_{\kappa\lambda\rho\sigma}+J^4_{\kappa\lambda\rho\sigma})\bigg]
\nonumber\end{eqnarray}}\end{enumerate}
\section{Propagators\label{propagator}}
Here we summarize the quark propagator in the presence of the external
gauge field in the fixed point gauge\cite{Niv84}.
{\scriptsize
\begin{eqnarray}
iS(p) & \equiv & \int d^4x e^{ipx}\,iS(x,0) \nonumber\\
&=&iS^{(0)}(p)+\int d^4x e^{ipx}\,g\int d^4 z iS^{(0)} (x-z) i\not\!A (z) iS^{(0)} (z) \nonumber\\
&&+\int d^4x e^{ipx}\,g^2 \int d^4z^\prime d^4z
iS^{(0)} (x-z^\prime) i\not\!A (z^\prime) iS^{(0)} (z^\prime-z) i\not\!A (z) iS^{(0)} (z) \nonumber\\
&&+\cdots
\end{eqnarray}}
and
\begin{eqnarray}
i{\tilde S}(p) & \equiv & \int d^4x e^{-ipx}\,iS(0,x).
\end{eqnarray}
In the fixed point gauge, we write the field in terms of
covariant operators,
\begin{eqnarray}
A_\mu(x)&=&\frac{1}{2\cdot 0!}x_\rho G_{\rho\mu}(0)+
\frac{1}{3\cdot 1!}x_\alpha x_\rho \left(D_\alpha G_{\rho\mu}(0)\right)\nonumber\\
&&+
\frac{1}{4\cdot 2!}x_\alpha x_\beta x_\rho \left(D_\alpha D_\beta G_{\rho\mu}(0)\right)+
\cdots
\end{eqnarray}

Collecting terms, we can write the full propagator in terms of
gauge covariant fields.
A few symbols are used for convenience:

\begin{eqnarray}
{\hat x}_\alpha\equiv -i\frac{\stackrel{\rightarrow}{\partial}}{\partial p},
~~~
\stackrel{\leftarrow}{x}_\alpha \equiv i\frac{\stackrel{\leftarrow}{\partial}}{\partial p},
\end{eqnarray}
and
\begin{eqnarray}
&&\{\alpha,\,\beta\}=\frac{1}{\not\!p-m} \gamma_\alpha \frac{1}{\not\!p-m} \gamma_\beta \frac{1}{\not\!p-m}
\nonumber\\
&&\{\alpha,\,\beta,\,\sigma \}=
\frac{1}{\not\!p-m} \gamma_\alpha \frac{1}{\not\!p-m} \gamma_\beta \frac{1}{\not\!p-m}
\gamma_\sigma \frac{1}{\not\!p-m}\nonumber\\
&&~~~\vdots,
\end{eqnarray}
and
$P( \alpha,\beta,\gamma...)$ means sum of all possible permutations in $\alpha,\beta,
\gamma...$.

Here we simply list the propagators.
\begin{enumerate}

\item{$iS^{(0)} (p)=i{\tilde S}^{(0)} (p)$}
{\footnotesize\begin{eqnarray}
=\frac{i}{\not\!p-m}\nonumber
\end{eqnarray}}

\item{$iS_G (p)=i{\tilde S}_G (p)$}
 {\footnotesize\begin{eqnarray}
 &=&iS^{(0)}(p) ~i(\gamma_\beta \cdot \frac{1}{2} {\hat x}_\alpha G_{\alpha\beta})~
 iS^{(0)}(p) =\frac{1}{2}\, G_{\alpha\beta} \{\beta,\,\alpha\}\nonumber\\
 &=&iS^{(0)}(p) ~i(\gamma_\beta \cdot \frac{1}{2} \stackrel{\leftarrow}{x}_\alpha G_{\alpha\beta})~
 iS^{(0)}(p) =-\frac{1}{2}\, G_{\alpha\beta} \{\alpha,\,\beta\}
 \nonumber\end{eqnarray}}

\item{$iS_{G^2} (p)=i{\tilde S}_{G^2} (p)$}
 {\footnotesize\begin{eqnarray}
 &=&iS^{(0)}(p) ~i(\gamma_\beta \cdot \frac{1}{2} {\hat x}_\alpha G_{\alpha\beta})~
 iS^{(0)}(p) ~i(\gamma_\sigma \cdot \frac{1}{2} {\hat x}_\rho G_{\rho\sigma})~
 iS^{(0)}(p) \nonumber\\
 &=&-\frac{i}{4} G_{\alpha\beta}G_{\rho\sigma}\Big[ \{\beta,\alpha,\sigma,\rho \}+
 \{\beta, \sigma,P(\alpha, \rho)\}\Big]\nonumber\\
 &=&iS^{(0)}(p) ~i(\gamma_\beta \cdot \frac{1}{2} \stackrel{\leftarrow}{x}_\alpha G_{\alpha\beta})~
 iS^{(0)}(p) ~i(\gamma_\sigma \cdot \frac{1}{2} \stackrel{\leftarrow}{x}_\rho G_{\rho\sigma})~
 iS^{(0)}(p) \nonumber\\
 &=&-\frac{i}{4} G_{\alpha\beta}G_{\rho\sigma}\Big[ \{\alpha,\beta,\rho,\sigma \}+
 \{P(\alpha, \rho),\beta, \sigma\}\Big]
 \nonumber\end{eqnarray}}

\item As for $G^3$ part,

\begin{itemize}
\item{$iS_{G^3} (p)$}
 {\scriptsize\begin{eqnarray}
 &=&iS^{(0)}(p) ~i(\gamma_\beta \frac{1}{2} {\hat x}_\alpha G_{\alpha\beta})~
 iS^{(0)}(p) ~i(\gamma_\sigma \frac{1}{2} {\hat x}_\rho G_{\rho\sigma})~
 iS^{(0)}(p) ~i(\gamma_\lambda \frac{1}{2} {\hat x}_\kappa G_{\kappa\lambda})~
 iS^{(0)}(p) \nonumber\\
 &=&-\frac{1}{8}
 G_{\alpha\beta}G_{\rho\sigma}G_{\kappa\lambda}\Big[\{\beta,\alpha,\sigma,\rho,\lambda,\kappa\}
 +\{\beta,\alpha,\sigma,\lambda,P(\rho,\kappa)\} \nonumber\\
 &&~~~~~~~+\{\beta,\sigma,P(\alpha,\rho),\lambda,\kappa\}
 +\{\beta,\sigma,\rho,\lambda,P(\alpha,\kappa)\}\nonumber\\
 &&~~~~~~~+\{\beta,\sigma,\alpha,\lambda,P(\rho,\kappa)\}
 +\{\beta,\sigma,\lambda,P(\alpha,\rho,\kappa)\}
 \Big] \nonumber
 \end{eqnarray}}

\item{$i{\tilde S}_{G^3} (p)$}
 {\scriptsize\begin{eqnarray}
 &=&iS^{(0)}(p) ~i(\gamma_\beta \frac{1}{2} {\stackrel{\leftarrow}{x}}_\alpha G_{\alpha\beta})~
 iS^{(0)}(p) ~i(\gamma_\sigma \frac{1}{2} {\stackrel{\leftarrow}{x}}_\rho G_{\rho\sigma})~
 iS^{(0)}(p) ~i(\gamma_\lambda \frac{1}{2} {\stackrel{\leftarrow}{x}}_\kappa G_{\kappa\lambda})~
 iS^{(0)}(p) \nonumber\\
 &=&\frac{1}{8}
 G_{\alpha\beta}G_{\rho\sigma}G_{\kappa\lambda}\Big[\{\alpha,\beta,\rho,\sigma,\kappa,\lambda\}
 +\{P(\alpha,\rho), \beta,\sigma,\kappa,\lambda\}\nonumber\\
 &&~~~~~~~+\{\alpha,\beta,P(\kappa,\rho),\sigma,\lambda\}
 +\{ P(\alpha,\kappa,\rho),\beta,\sigma,\lambda\}\nonumber\\
 &&~~~~~~~+\{P(\alpha,\rho),\beta,\kappa,\sigma,\lambda\}
 +\{P(\alpha,\kappa),\beta,\rho,\sigma,\lambda\} \Big] \nonumber
 \end{eqnarray}}
 \end{itemize}

\item As for $DG$ part,
\begin{itemize}
\item{$iS_{DG} (p)$}
 {\footnotesize\begin{eqnarray}
 &=&iS^{(0)}(p)~i(\gamma_\sigma \cdot \frac{1}{3} {\hat x}_\alpha {\hat x}_\beta D_\alpha
 G_{\beta\sigma})~iS^{(0)}(p)\nonumber\\
 &=&\frac{i}{3} D_\alpha G_{\beta\sigma} \{\sigma, P(\alpha,\beta)\}
 \nonumber\end{eqnarray}}

\item{$i{\tilde S}_{DG} (p)$}
 {\footnotesize\begin{eqnarray}
 &=&iS^{(0)}(p)~i(\gamma_\sigma \cdot \frac{1}{3} {\stackrel{\leftarrow}{x}}_\alpha
 {\stackrel{\leftarrow}{x}}_\beta D_\alpha
 G_{\beta\sigma})~iS^{(0)}(p)\nonumber\\
 &=&\frac{i}{3} D_\alpha G_{\beta\sigma} \{P(\alpha,\beta),\sigma\}
 \nonumber\end{eqnarray}}
\end{itemize}

\item As for $D^2G$ part,
\begin{itemize}
\item{$iS_{D^2G} (p)$}
 {\footnotesize\begin{eqnarray}
 &=&iS^{(0)}(p)~i(\gamma_\sigma \cdot \frac{1}{8} {\hat x}_\alpha{\hat x}_\beta{\hat x}_\nu
 D_\alpha D_\beta G_{\nu\sigma})~iS^{(0)}(p) \nonumber\\
 &=&-\frac{1}{8} D_\alpha D_\beta G_{\nu\sigma} \{\sigma,P(\alpha,\beta,\nu)\}
 \nonumber\end{eqnarray}}

\item{$i{\tilde S}_{D^2G} (p)$}
 {\footnotesize\begin{eqnarray}
 &=&iS^{(0)}(p)~i(\gamma_\sigma \cdot \frac{1}{8} {\stackrel{\leftarrow}{x}}_\alpha
 {\stackrel{\leftarrow}{x}}_\beta {\stackrel{\leftarrow}{x}}_\nu
 D_\alpha D_\beta G_{\nu\sigma})~iS^{(0)}(p) \nonumber\\
 &=&+\frac{1}{8} D_\alpha D_\beta G_{\nu\sigma} \{P(\alpha,\beta,\nu),\sigma\}
 \nonumber\end{eqnarray}}
\end{itemize}

\item As for $(DG)^2$ part,

\begin{itemize}
\item{$iS_{(DG)^2} (p)$}
 {\scriptsize\begin{eqnarray}
 &=&iS^{(0)}(p)~i(\gamma_\kappa \cdot\frac{1}{3} {\hat x}_\alpha{\hat x}_\beta D_\alpha
 G_{\beta\kappa})~
 iS^{(0)}(p)~i(\gamma_\lambda \cdot\frac{1}{3} {\hat x}_\rho{\hat x}_\sigma D_\rho
 G_{\sigma\lambda})~iS^{(0)}(p)
\nonumber\\
 &=&\frac{i}{9} D_\alpha G_{\beta\kappa} D_\rho
 G_{\sigma\lambda}\Big[
 \{\kappa, P(\alpha,\beta), \lambda, P(\rho,\sigma)\}
\nonumber\\
&&~~~+\{\kappa, \alpha, \lambda, P(\beta,\rho,\sigma)\}
 +\{\kappa, \beta, \lambda, P(\alpha,\rho,\sigma)\}
 +\{\kappa,\lambda,P(\alpha,\beta,\rho,\sigma )\} \Big]
 \nonumber\end{eqnarray}}

\item{$i{\tilde S}_{(DG)^2} (p)$}
 {\scriptsize\begin{eqnarray}
 &=&iS^{(0)}(p)~i(\gamma_\kappa \cdot\frac{1}{3} {\stackrel{\leftarrow}{x}}_\alpha
 {\stackrel{\leftarrow}{x}}_\beta D_\alpha
 G_{\beta\kappa})~
 iS^{(0)}(p)~i(\gamma_\lambda \cdot\frac{1}{3} {\stackrel{\leftarrow}{x}}_\rho
 {\stackrel{\leftarrow}{x}}_\sigma D_\rho
 G_{\sigma\lambda})~iS^{(0)}(p)
\nonumber\\
 &=&\frac{i}{9} D_\alpha G_{\beta\kappa} D_\rho
 G_{\sigma\lambda}\Big[
 \{P(\alpha,\beta),\kappa, P(\rho,\sigma),  \lambda\}
\nonumber\\
&&~~~+\{P(\alpha,\beta,\rho), \kappa,\sigma,\lambda\}
 +\{P(\alpha,\beta,\sigma), \kappa,\rho,\lambda\}
 +\{P(\alpha,\beta,\rho,\sigma ),\kappa,\lambda\} \Big]
 \nonumber\end{eqnarray}}
\end{itemize}

\item As for $GD^2G$ part,

\begin{itemize}
\item{$iS_{GD^2G} (p)$}
 {\footnotesize\begin{eqnarray}
 &=&iS^{(0)}(p)~i(\gamma_\beta \cdot \frac{1}{2} {\hat x}_\alpha G_{\alpha\beta})~
 iS^{(0)}(p)~i(\gamma_\sigma \cdot \frac{1}{8} {\hat x}_\rho{\hat x}_\kappa
 {\hat x}_\lambda D_\rho D_\kappa
 G_{\lambda\sigma})~iS^{(0)}\nonumber\\
 &=&\frac{i}{16} G_{\alpha\beta}D_\rho D_\kappa G_{\lambda\sigma}
 \Big[\{\beta, \alpha, \sigma, P(\rho, \kappa, \lambda) \}
 +\{\beta, \sigma, P(\alpha,\rho,\kappa,\lambda)\}\Big]
 \nonumber\end{eqnarray}}

\item{$i{\tilde S}_{GD^2G} (p)$}
 {\footnotesize\begin{eqnarray}
 &=&iS^{(0)}(p)~i(\gamma_\beta \cdot \frac{1}{2} {\stackrel{\leftarrow}{x}}_\alpha G_{\alpha\beta})~
 iS^{(0)}(p)~i(\gamma_\sigma \cdot \frac{1}{8} {\stackrel{\leftarrow}{x}}_\rho
 {\stackrel{\leftarrow}{x}}_\kappa
 {\stackrel{\leftarrow}{x}}_\lambda D_\rho D_\kappa
 G_{\lambda\sigma})~iS^{(0)}\nonumber\\
 &=&\frac{i}{16} G_{\alpha\beta}D_\rho D_\kappa G_{\lambda\sigma}
 \Big[\{ \alpha,\beta, P(\rho,\kappa,\lambda),\sigma \}+
 \{P(\alpha,\rho),\beta,P(\kappa,\lambda),\sigma\}
 \nonumber\\
 &&~~~~~~~+\{P(\alpha,\kappa),\beta,P(\rho,\lambda),\sigma \}
 +\{P(\alpha,\lambda),\beta,P(\kappa,\rho),\sigma \}
 \nonumber\\
 &&~~~~~~~+\{P(\alpha,\kappa,\lambda),\beta,\rho,\sigma\}
 +\{P(\alpha,\rho,\lambda),\beta,\kappa,\sigma\}
 \nonumber\\
 &&~~~~~~~
 +\{P(\alpha,\kappa,\rho),\beta,\lambda,\sigma\}
 +\{P(\alpha,\rho,\kappa,\lambda),\beta,\sigma\}
 \Big]
 \nonumber\end{eqnarray}}
\end{itemize}

\item As for $D^2GG$ part,

\begin{itemize}
\item{$iS_{D^2GG} (p)$}
 {\footnotesize\begin{eqnarray}
 &=&iS^{(0)}(p)~i(\gamma_\sigma \cdot \frac{1}{8} {\hat x}_\rho{\hat x}_\kappa
 {\hat x}_\lambda D_\rho D_\kappa G_{\lambda\sigma})~iS^{(0)}(p)
 ~i(\gamma_\beta \cdot \frac{1}{2} {\hat x}_\alpha G_{\alpha\beta})~
 iS^{(0)}(p)\nonumber\\
 &=&\frac{i}{16} D_\rho D_\kappa G_{\lambda\sigma}
 G_{\alpha\beta}\Big[
 \{\sigma, P(\rho,\kappa,\lambda),\beta,\alpha\}
 +\{\sigma, P(\kappa,\lambda),\beta, P(\alpha,\rho)\}\nonumber\\
 &&~~~~~~~+\{\sigma, \rho,\beta,P(\kappa,\lambda,\alpha)\}
 +\{\sigma, \beta,P(\rho,\kappa,\lambda,\alpha)\}\nonumber\\
 &&~~~~~~~+\{\sigma, P(\kappa,\rho), \beta, P(\alpha,\lambda)\}
 +\{\sigma, \kappa, \beta, P(\alpha,\lambda,\rho)\}\nonumber\\
 &&~~~~~~~+\{\sigma, P(\lambda,\rho),\beta,P(\alpha,\kappa)\}
 +\{\sigma, \lambda, \beta, P(\rho,\alpha,\kappa)\}\Big]\nonumber\\
 \nonumber\end{eqnarray}}

\item{$i{\tilde S}_{D^2GG} (p)$}
 {\footnotesize\begin{eqnarray}
 &=&iS^{(0)}(p)~i(\gamma_\sigma \cdot \frac{1}{8} {\stackrel{\leftarrow}{x}}_\rho
 {\stackrel{\leftarrow}{x}}_\kappa
 {\stackrel{\leftarrow}{x}}_\lambda D_\rho D_\kappa G_{\lambda\sigma})~iS^{(0)}(p)
 ~i(\gamma_\beta \cdot \frac{1}{2} {\stackrel{\leftarrow}{x}}_\alpha G_{\alpha\beta})~
 iS^{(0)}(p)\nonumber\\
 &=&\frac{i}{16} D_\rho D_\kappa G_{\lambda\sigma}
 G_{\alpha\beta}\Big[
 \{P(\rho,\kappa,\lambda),\sigma,\alpha,\beta\}+
 \{P(\rho,\kappa,\lambda,\alpha),\sigma,\beta\}\Big]\nonumber\\
 \nonumber\end{eqnarray}}
\end{itemize}
\end{enumerate}
\section{spin structure of operators of dimension 6}
In computing the Wilson coefficients we used the following reduction of the
Lorentz indices to the spin-2 operators.
\begin{eqnarray}
f^{abc} G^a_{\mu\nu} G^b_{\alpha\beta} G^c_{\rho\sigma}
&=&A_{\mu\alpha} c_{\nu\rho\beta\sigma} -A_{\mu\beta} c_{\nu\rho\alpha\sigma}
-A_{\nu\alpha} c_{\mu\rho\beta\sigma} +A_{\nu\beta} c_{\mu\rho\alpha\sigma}
\nonumber\\
& & -A_{\mu\rho} c_{\nu\alpha\sigma\beta} +A_{\mu\sigma}c_{\nu\alpha\rho\beta}
+A_{\nu\rho} c_{\mu\alpha\sigma\beta} -A_{\nu\sigma}c_{\mu\alpha\rho\beta}
\nonumber\\
&&+A_{\alpha\rho} c_{\beta\mu\sigma\nu}-A_{\alpha\sigma} c_{\beta\mu\rho\nu}
-A_{\beta\rho} c_{\alpha\mu\sigma\nu} +A_{\beta\sigma} c_{\alpha\mu\rho\nu}
\\[12pt]
G^a_{\mu_1 \nu_1} G^a_{\mu_2 \nu_2;\alpha\beta}
&=&K_{\alpha\beta} c_{\mu_1 \mu_2 \nu_1 \nu_2} \nonumber\\
&&+P_{\beta\mu_1}c_{\alpha\mu_2 \nu_1\nu_2}-P_{\beta\nu_1}c_{\alpha\mu_2 \mu_1\nu_2}\nonumber\\
&&~+J_{\alpha\mu_2}c_{\beta\mu_1 \nu_2 \nu_1}-J_{\alpha\nu_2}c_{\beta\mu_1 \mu_2 \nu_1}\nonumber\\
&&+Q_{\beta\mu_2}c_{\alpha\mu_1 \nu_2 \nu_1} -Q_{\beta\nu_2}c_{\alpha\mu_1 \mu_2 \nu_1}\nonumber\\
&&~+W_{\alpha\mu_1}c_{\beta\mu_2 \nu_1 \nu_2}-W_{\alpha\nu_1} c_{\beta\mu_2 \mu_1 \nu_2}\nonumber\\
&&+L_{\mu_1 \mu_2}d_{\beta\nu_1 \alpha\nu_2}-L_{\nu_1\mu_2} d_{\beta\mu_1 \alpha\nu_2}\nonumber\\
&&~-L_{\mu_1 \nu_2} d_{\beta\nu_1 \alpha\mu_2}+L_{\nu_1 \nu_2} d_{\beta\mu_1 \alpha\mu_2}\nonumber\\
&&+M_{\mu_1 \mu_2}d_{\beta\alpha\nu_1\nu_2} -M_{\nu_1\mu_2} d_{\alpha\beta\mu_1 \nu_2}\nonumber\\
&&~-M_{\mu_1 \nu_2} d_{\alpha\beta\nu_1\mu_2} +M_{\nu_1\nu_2}d_{\alpha\beta\mu_1 \mu_2}\nonumber\\
&&+T_{\mu_1 \mu_2}d_{\beta\nu_2 \alpha\nu_1} -T_{\nu_1\mu_2} d_{\beta\nu_2 \alpha\mu_1}\nonumber\\
&&~-T_{\mu_1 \nu_2} d_{\beta\mu_2 \alpha\nu_1} +T_{\nu_1 \nu_2} d_{\beta\mu_2 \alpha \mu_1},
\end{eqnarray}
where
\begin{eqnarray}
c_{\alpha\beta\mu\nu}&\equiv&g_{\alpha\beta} g_{\mu\nu}-g_{\alpha\nu} g_{\mu\beta} ,
\nonumber\\
d_{\alpha\beta\mu\nu}&\equiv&g_{\alpha\beta} g_{\mu\nu}
\end{eqnarray}
and,
if we take 3 operators, $ddgg1\equiv G^a_{\kappa\lambda} G^a_{\kappa\lambda;\mu\nu},
~ddgg2\equiv-g^2j^a_\mu j^a_\nu,$ and
$ddgg3\equiv G^a_{\mu\kappa} G^a_{\kappa\lambda;\lambda\nu}$ as our basis, we get
\begin{eqnarray}
&&A=(ddgg1+2ddgg3)/4, \nonumber\\
&&K=2P=2J=(13ddgg1+ddgg2+19ddgg3)/80, \nonumber\\
&&Q=W=(-27ddgg1+ddgg2-61ddgg3)/160, \nonumber\\
&&L=(4ddgg1+3ddgg2+7ddgg3)/40, \nonumber\\
&&M=(29ddgg1 +13ddgg2+47ddgg3)/160, \nonumber\\
&&T=(-6ddgg1 +3ddgg2-13ddgg3)/40.
\end{eqnarray}

As for the spin 4 part, we have a simpler reduction:
\begin{eqnarray}
G^a_{\mu_1 \nu_1} G^a_{\mu_2 \nu_2;\alpha\beta}
&=&W_{\alpha\beta\mu_1\mu_2} g_{\nu_1 \nu_2}+
W_{\alpha\beta\nu_1 \nu_2} g_{\mu_1\mu_2}\nonumber\\
&&-W_{\alpha\beta\mu_1\nu_2} g_{\nu_1 \mu_2}-
W_{\alpha\beta\nu_1 \mu_2} g_{\mu_1\nu_2},
\end{eqnarray}
where
\begin{eqnarray}
W_{\alpha\beta\mu_1\mu_2}=\frac{1}{2}
G^a_{\mu_1 \kappa} G^a_{\mu_2 \kappa;\alpha\beta}.
\end{eqnarray}


\section{Integrations with respect to Feynman Parameter}
In general, after Feynman integration, one can write the
polarizations in terms of linear sums of $J$'s defined in eq.(\ref{defJ}).
\begin{eqnarray}
\label{Int}
\Pi(q)=Q^{-a}\left[\, a_0J_0 +a_1J_1 +a_2J_2 +\cdots\,\, +a_kJ_k\right].
\end{eqnarray}

To reduce the polarization function into this final form, we use
the following steps and identities.

After the Feynman integral, the polarization function will be a sum of
 $I{mn}$'s:
\begin{eqnarray}
I^{mn}_N (Q^2,m^2)\equiv\int^1_0 dx\frac{x^n (1-x)^m}{\left[ m^2+Q^2 x(1-x) \right]^N}
=I^{nm}_N (Q^2,m^2)
\end{eqnarray}
We then follow the following steps.
\begin{enumerate}
\item  $I^{mn}$ can be expressed in terms of $I^n_N \equiv I^{nn}_N $
using the following identity.
{\scriptsize\begin{eqnarray}
x^n +(1-x)^n &=& 1-\Big\{ ~_n C_1 x^{n-1} (1-x) +_nC_2 x^{n-2} (1-x)^2 +\cdots +_nC_k x^{n-k} (1-x)^k \nonumber\\
&&+\cdots+_nC_{n-1} x (1-x)^{n-1}  \Big\}\nonumber\\
&=& 1-_n\!C_1 \{ x^{n-1} (1-x) +x(1-x)^{n-1} \}
\nonumber\\
&&-_n\!C_2  \{ x^{n-2} (1-x)^2 +x^2 (1-x)^{n-2} \}-\cdots \nonumber\\
&=& 1-_n\!C_1 ~x(1-x) \Big\{ x^{n-2} +(1-x)^{n-2}\Big\}\nonumber\\
&&- _n\!C_2 ~x^2(1-x)^2 \Big\{x^{n-4} +(1-x)^{n-4}\Big\}-\cdots \nonumber
\end{eqnarray}}
\item Then we can reduce $I^n_N$ to $I^0_N \equiv I_N$ using
$ I^n_N =\frac{1}{Q^2} (I^{n-1}_{N-1} -m^2 I^{n-1}_N)
=\frac{1}{Q^2} I^{n-1}_{N-1}-\frac{1}{y} I^{n-1}_{N-1}$
\item We then introduce the dimensionless function
$J_N (y) = \left( \frac{Q^2}{y} \right)^N I_N$, where $y=Q^2/m^2$.
\item Finally, we use the recurrence relation.
\begin{eqnarray}
y\,J_N(y)=\frac{2}{N-1}+\frac{4N-6}{N-1}J_{N-1}(y)-4J_N(y)
\end{eqnarray}
\item We can also
integrate explicitly,
{\scriptsize\begin{eqnarray}
J_N(y)=\frac{(2N-3)!!}{(N-1)!} \left[
\left(\frac{s-1}{2s}\right)^N \sqrt{s} \log\frac{\sqrt{s}+1}{\sqrt{s}-1} +\sum^{N-1}_{k=1}
\frac{(k-1)!}{(2k-1)!!}\left( \frac{s-1}{2s}\right)^{N-k}
\right],
\end{eqnarray}}
where $s=1+4/y$.
\end{enumerate}

\section{An example: evaluation of  a diagram}

{\bf Diagram 1a}: We show the evaluation of the Feynman diagrams here.
The first diagram in Fig. 1 is taken as an example.
{\scriptsize
\begin{eqnarray}
&&i\int \frac{d^4 k}{(2\pi)^4}
{\rm Tr}\left[ \gamma_\mu S^{(0)}(k+q) \gamma_\nu {\tilde S}_{G^3}(k)\right]
\nonumber\\
&=&i\int \frac{d^4 k}{(2\pi)^4}
{\rm Tr}\bigg[ \gamma_\mu \frac{1}{\not\!k +\not\!q -m} \gamma_\nu \,
\left( \frac{-i}{8}\right)
 G^a_{\alpha\beta}G^b_{\rho\sigma}G^c_{\kappa\lambda} t^a t^b t^c
 \Big[
 \{\alpha,\beta,\rho,\sigma,\kappa,\lambda\}
 +\{P(\alpha,\rho), \beta,\sigma,\kappa,\lambda\}
 \nonumber\\ &&~~~
 +\{\alpha,\beta,P(\kappa,\rho),\sigma,\lambda\}
 +\{ P(\alpha,\kappa,\rho),\beta,\sigma,\lambda\}
 +\{P(\alpha,\rho),\beta,\kappa,\sigma,\lambda\}
 +\{P(\alpha,\kappa),\beta,\rho,\sigma,\lambda\} \Big]
\bigg]\nonumber\\
&=&\frac{1}{8}\frac{i}{4}\left( f^{abc}\,G^a_{\alpha\beta}G^b_{\rho\sigma}G^c_{\kappa\lambda} \right)
\int \frac{d^4 k}{(2\pi)^4}
\frac{1}{(k+q)^2-m^2}
{\rm Tr}\bigg[ \gamma_\mu (\not\!k +\not\!q +m) \gamma_\nu \nonumber\\
 &&~~~\times \Big[\{\alpha,\beta,\rho,\sigma,\kappa,\lambda\}
 +\{P(\alpha,\rho), \beta,\sigma,\kappa,\lambda\}
 +\{\alpha,\beta,P(\kappa,\rho),\sigma,\lambda\}\nonumber\\
 &&~~~+\{ P(\alpha,\kappa,\rho),\beta,\sigma,\lambda\}
 +\{P(\alpha,\rho),\beta,\kappa,\sigma,\lambda\}
 +\{P(\alpha,\kappa),\beta,\rho,\sigma,\lambda\} \Big]
\bigg]\nonumber\\
&=&\frac{1}{8}\frac{i}{4}
\Big(\,A_{\mu\alpha} c_{\nu\rho\beta\sigma} -A_{\mu\beta} c_{\nu\rho\alpha\sigma}
-A_{\nu\alpha} c_{\mu\rho\beta\sigma} +A_{\nu\beta} c_{\mu\rho\alpha\sigma}
-A_{\mu\rho} c_{\nu\alpha\sigma\beta} +A_{\mu\sigma}c_{\nu\alpha\rho\beta}
\nonumber\\
&&~~~~+A_{\nu\rho} c_{\mu\alpha\sigma\beta} -A_{\nu\sigma}c_{\mu\alpha\rho\beta}
+A_{\alpha\rho} c_{\beta\mu\sigma\nu}-A_{\alpha\sigma} c_{\beta\mu\rho\nu}
-A_{\beta\rho} c_{\alpha\mu\sigma\nu} +A_{\beta\sigma} c_{\alpha\mu\rho\nu}\Big)
\nonumber\\
&&\times\int \frac{d^4 k}{(2\pi)^4}
\frac{1}{(k+q)^2-m^2}\frac{1}{(k^2-m^2)^7}
{\rm Tr}\bigg[ \gamma_\mu (\not\!k +\not\!q +m) \gamma_\nu \nonumber\\
 &&\times \Big\{ (\not\!k+m)\gamma_\alpha(\not\!k+m)\gamma_\beta(\not\!k+m)
\gamma_\rho(\not\!k+m)\gamma_\sigma(\not\!k+m)\gamma_\kappa
 (\not\!k+m)\gamma_\lambda(\not\!k+m)+\cdots \Big\}  \bigg]\nonumber\\
&\equiv&i\int \frac{d^4 k}{(2\pi)^4} \frac{1}{(k+q)^2-m^2}\frac{1}{(k^2-m^2)^7} f_{1a}(k)\nonumber\\
&=&i\int \frac{d^4 k}{(2\pi)^4} \left[
\int^1_0 dx \frac{x^6}{[ \{k+(1-x)q\}^2 -s^2]^8} \frac{\Gamma(8)}{\Gamma(7)}
\right] f_{1a}(k)\nonumber\\
&=&\frac{i}{8}\int^1_0 dx \int \frac{d^4 l}{(2\pi)^4} \frac{1}{[ l^2 -s^2]^8}
\frac{\Gamma(8)}{\Gamma(7)}
\Bigg[ \Big\{ -\frac{28}{3}l^8 +\left( \frac{116}{3}m^2 -\frac{220}{3} q^2t^2 \right)l^6 \nonumber\\
&&+
\left( -\frac{148}{3} m^4 +\frac{416}{3}m^2q^2t^2 -\frac{268}{3} q^4t^4\right) l^4
+
\left(20m^6 -60m^4q^2t^2+60m^2q^4t^4-20q^6t^6 \right)l^2 \Big\}A_{\mu\nu} \nonumber\\
&&+\Big\{ (-640t^3+896t^4)l^4 +(640m^2t^3-768m^2t^4 )l^2\nonumber\\
&& -128m^4(t^3-t^4)+256m^2q^2(t^5-t^6)-128q^4(t^7-t^8)\Big\}
q_\mu q_\nu q_\alpha q_\beta A_{\alpha\beta} \nonumber\\
&&+\Big\{ -\frac{140}{3} tl^6 +
\left(\frac{464}{3}m^2t+\frac{400}{3}m^2t^2+\frac{80}{3}q^2t^3 \right)l^4 \nonumber\\
&&+
(-148m^4t-112m^4t^2+96m^2q^2t^3+296m^2q^2t^4+116q^4t^5-184q^4t^6 )l^2\nonumber\\
&& +(40t+24t^2)m^6-(56q^2t^3-72q^2t^4)m^4+(-8q^4t^5 +72q^4t^6)m^2
+24q^6(t^7-t^8)\Big\}
g_{\mu\nu} q_\alpha q_\beta A_{\alpha\beta}
\nonumber\\
&&+\Big\{\left( \frac{140}{3}t -40t^2\right)l^6+
(-\frac{464}{3}m^2t+\frac{560}{3}m^2t^2)l^4\nonumber\\
&&+
(148m^4t-176m^4t^2-416m^2q^2t^3+472m^2q^2t^4+268q^4t^5-296q^4t^6)l^2\nonumber\\
&&-
40m^6(t-t^2)+120m^4q^2(t^3-t^4)-120m^2q^4(t^5-t^6)+40q^6(t^7-t^8)
\Big\}
(q_\mu q_\alpha A_{\alpha\nu} +q_\nu q_\alpha A_{\alpha\mu} )\Bigg]\nonumber\\
&=&
A_{\mu\nu}\frac{\pi^2}{Q^4}\left(
\frac{7}{3}-y+(\frac{1}{3}+\frac{1}{6}y )J_1-\frac{8}{3}J_2\right)
\nonumber\\&&+
g_{\mu\nu} q_\alpha q_\beta A_{\alpha\beta}\frac{\pi^2}{Q^6}\left(
\frac{5}{3} +y-(2+y)J_1-9J_2+\frac{52}{3}J_3-8J_4\right)
\nonumber\\&&+
q_\mu q_\nu q_\alpha q_\beta A_{\alpha\beta}\frac{\pi^2}{Q^8}\left(
\frac{68}{3}-(8+4y)J_1-36J_2+\frac{88}{3}J_3-8J_4\right)
\nonumber\\&&+
(q_\mu q_\alpha A_{\alpha\nu}+q_\nu q_\alpha)A_{\alpha\mu}
\frac{\pi^2}{Q^6}\left(
\frac{23}{3}-y-(\frac{2}{3}+\frac{1}{3}y)J_1
-\frac{41}{3}J_2+\frac{20}{3}J_3\right)
\end{eqnarray}}

\section{Moments in terms of Hypergeometric Functions $_2F_1$\label{hyp}}
In order to obtain moments written in terms of hypergeometric functions $_2F_1$ from
polarizations eq.(\ref{Int}),
we expand the  polarizations in terms of $Q^2/m^2$.
Here we divide $a_0$ and $a_1$ into $a_0=d_1+d_2y$ and $a_1=c_1+c_2y$.
The other higher coefficients, $a_2,a_3,\cdots,a_k$ don't contain $y$, where
$y=Q^2/m^2$.
{\scriptsize\begin{eqnarray}
\Pi(q)&=&Q^{-a}\bigg[\, a_0 +(c_1+c_2y)\int^1_0\frac{dx}{[1+x(1-x)Q^2/m^2]}
+a_2\int^1_0\frac{dx}{[1+x(1-x)Q^2/m^2]^2} +\nonumber\\
&&~~~\vdots\nonumber\\
&&
 +a_k\int^1_0\frac{dx}{[1+x(1-x)Q^2/m^2]^k} \bigg]\nonumber\\
&=&Q^{-a}\bigg[\, a_0 +(c_1+c_2y)\int^1_0dx \sum^{\infty}_{j=0} \left\{-x(1-x)\right\}^j
\frac{Q^{2j}}{m^{2j}} \nonumber\\
&&
+a_2\int^1_0dx \sum^{\infty}_{j=0} \left\{-x(1-x)\right\}^j
\frac{(1+j)!}{j!}  \frac{Q^{2j}}{m^{2j}} +\nonumber\\
&&~~~\vdots\nonumber\\
&&+a_k \int^1_0dx \sum^{\infty}_{j=0} \left\{-x(1-x)\right\}^j
\frac{(k-1+j)!}{(k-1)!j!}
\frac{Q^{2j}}{m^{2j}}\bigg]\nonumber\\
&=&Q^{-a}\sum^{\infty}_{j=1}\Bigg[\, c_1\int^1_0dx  (-1)^j x^j(1-x)^j
\frac{Q^{2j}}{m^{2j}}
+a_2\int^1_0dx (-1)^j x^j(1-x)^j
(1+j) \frac{Q^{2j}}{m^{2j}} +\nonumber\\
&&~~~\vdots\nonumber\\
&&+a_k \int^1_0dx  (-1)^j x^j(1-x)^j
\frac{(k-1+j)!}{(k-1)!j!}
\frac{Q^{2j}}{m^{2j}}
-c_2 \int^1_0dx  (-1)^j x^{j-1}(1-x)^{j-1}
\frac{Q^{2j}}{m^{2j}}\Bigg]\nonumber\\
&=&\sum^\infty_{j=a/2} (-)^{a/2-1} \int^1_0 dx \left\{ x(1-x) \right\}^{j-1}
\frac{(-Q^2)^{j-a/2}}{m^{2j}}\bigg[
-x(1-x)\Big\{ c_1 +a_2(1+j)
\nonumber\\&&
+\frac{a_3}{2!}(1+j)(2+j)+\cdots\,\,
+\frac{a_k}{(k-1)!}\prod^{k-1}_{l=1}(l+j)\Big\}+c_2 \bigg]
\end{eqnarray}}

Now the $n$'s  moment $M_n$  can be obtained from above by
differentiation.
{\scriptsize\begin{eqnarray}
M_n(Q_0) & = & \frac{1}{n!} \left( -\frac{d}{dQ^2}\right)^n \Pi(Q^2)
\nonumber\\
&=&(-)^{a/2}\frac{1}{(4m^2)^{a/2+n}}\frac{1}{(1+\xi)^{n+1}}\bigg[\,c_1d_1+c_2 d_2
\nonumber\\
&&+\frac{1}{2}\Bigg\{
a_2 \frac{\Gamma\left(\frac{1}{2}\right)\Gamma\left(n+\frac{a}{2}+2\right)}
{\Gamma\left(n+\frac{a+3}{2}\right)}\, _2F_1(n+1,-\frac{1}{2},n+\frac{a+3}{2}; \frac{\xi}{1+\xi})
\nonumber\\
&&+\frac{a_3}{2!}
\frac{\Gamma\left(\frac{1}{2}\right)\Gamma\left(n+\frac{a}{2}+3\right)}
{\Gamma\left(n+\frac{a+3}{2}\right)}\, _2F_1(n+1,-\frac{3}{2},n+\frac{a+3}{2}; \frac{\xi}{1+\xi})
\nonumber\\&&
+\frac{a_4}{3!}
\frac{\Gamma\left(\frac{1}{2}\right)\Gamma\left(n+\frac{a}{2}+4\right)}
{\Gamma\left(n+\frac{a+3}{2}\right)}\, _2F_1(n+1,-\frac{5}{2},n+\frac{a+3}{2}; \frac{\xi}{1+\xi})
\nonumber\\
&&+\cdots
\nonumber\\
&&+\frac{a_k}{(k-1)!}
\frac{\Gamma\left(\frac{1}{2}\right)\Gamma\left(n+\frac{a}{2}+k\right)}
{\Gamma\left(n+\frac{a+3}{2}\right)}~ _2F_1(n+1,-k+\frac{3}{2},n+\frac{a+3}{2}; \frac{\xi}{1+\xi})
\Bigg\}
\nonumber\\[15pt]
&=&
\frac{1}{(4m^2)^{a/2+n}}\frac{1}{(1+\xi)^{n+1}}
\frac{(-)^{a/2}\Gamma(1/2)}{2\Gamma(n+\frac{a+3}{2})} \Gamma\left(n+\frac{a}{2}\right)
\left( \begin{array}{c} c_1,c_2,a_2,a_3,\cdots~,a_k\end{array}\right)
\nonumber\\ [20pt] &&
\times
\left( \begin{array}{cc}
(n+\frac{a}{2})/0!~_2F_1(n+1,\frac{1}{2},n+\frac{a+3}{2}; \rho )\\
 -2(2n+a+1)~_2F_1(n+1,\frac{1}{2},n+\frac{a+1}{2}; \rho )\\
 (n+\frac{a}{2})(n+\frac{a+2}{2})/1!
~_2F_1(n+1,-\frac{1}{2},n+\frac{a+3}{2}; \rho )\\
 (n+\frac{a}{2})(n+\frac{a+2}{2})(n+\frac{a+4}{2})/2!
~_2F_1(n+1,-\frac{3}{2},n+\frac{a+3}{2}; \rho ) \\
 \vdots \\
(n+\frac{a}{2})(n+\frac{a+2}{2})\cdots (n+\frac{a}{2}+k-1)/(k-1)!
~_2F_1(n+1,-k+\frac{3}{2},n+\frac{a+3}{2}; \rho )
\end{array}\right)  \nonumber \\
\end{eqnarray}}
where
\begin{eqnarray}
 \left( \begin{array}{c} d_1\\d_2 \end{array} \right)
= \left( \begin{array}{cc} \frac{1}{2}
\frac{\Gamma[1/2]\,\Gamma[n+(a+2)/2 ]}
{\Gamma[n+(a+3)/2 ]}\, _2F_1(n+1,\frac{1}{2},n+\frac{a+3}{2}; \frac{\xi}{1+\xi})
\\
-2\frac{\Gamma[1/2]\,\Gamma[n+a/2]}
{\Gamma[n+(a+1)/2]}\, _2F_1(n+1,\frac{1}{2},n+\frac{a+1}{2}; \frac{\xi}{1+\xi})
\end{array}\right).
\end{eqnarray}
As an example, when $\xi=0$ and $a=6$, we have,
{\scriptsize\begin{eqnarray}
M_n(Q_0^2,\xi & =&0)= \frac{(-1)}{(4m_c^2)^{n+3}}\Bigg[
c_1
\frac{1}{2}
\frac{\Gamma\left(\frac{1}{2}\right)\Gamma\left(n+4\right)}
{\Gamma\left(n+\frac{9}{2}\right)} +
c_2
(-2)
\frac{\Gamma\left(\frac{1}{2}\right)\Gamma(n+3)\left(n+\frac{7}{2}\right)}
{\Gamma\left(n+\frac{9}{2}\right)} +
\nonumber\\
&&+\frac{1}{2}\bigg\{
a_2
\frac{\Gamma\left(\frac{1}{2}\right)\Gamma\left(n+5\right)}
{\Gamma\left(n+\frac{9}{2}\right)} +
\frac{a_3}{2!}
\frac{\Gamma\left(\frac{1}{2}\right)\Gamma\left(n+6\right)}
{\Gamma\left(n+\frac{9}{2}\right)} +\cdots\,
+\frac{a_k}{(k-1)!}
\frac{\Gamma\left(\frac{1}{2}\right)\Gamma\left(n+k+3\right)}
{\Gamma\left(n+\frac{9}{2}\right)}
\bigg\}\Bigg]\nonumber\\
&=&\frac{(-1)}{(4m_c^2)^{n+3}}\frac{2^{n+3}(n+2)!}{(2n+7)!!}
\left( \begin{array}{c} c_1,c_2,a_2,a_3,\cdots~,a_k\end{array}\right)
\left( \begin{array}{cc}
(n+3)/0! \\ -2(2n+7) \\ (n+3)(n+4)/1! \\ (n+3)(n+4)(n+5)/2! \\ \vdots \\
\frac{(n+2+k)!}{(n+2)!} /(k-1)!
\end{array}\right)\nonumber\\
&=& \frac{M^0_n}{(4m_c^2)^3}\left[ \frac{2^5 \pi^2}{3}\frac{-n(n+2)}{(2n+5)(2n+7)}
\right]
\left( \begin{array}{c} c_1,c_2,a_2,\cdots~,a_k\end{array}\right)
\left( \begin{array}{cc}
(n+3)/0! \\ -2(2n+7) \\ (n+3)(n+4)/1! \\ (n+3)(n+4)(n+5)/2! \\ \vdots \\
\frac{(n+2+k)!}{(n+2)!} /(k-1)!
\end{array}\right), \nonumber \\
\end{eqnarray}}
where
{\scriptsize$
M^0_n=3\times 2^n(n+1)(n-1)!/(4\pi^2 (2n+3)!!(4m_c^2)^n)=A_n(\xi=0).
$}

\section{Consistency check}

As discussed in the text, there are few checks we can perform to
confirm our calculation.  The first is the current conservation, which we
checked explicitly.  The second is the regularity at $Q^2=0$.
This can be checked from the following equation.

{\scriptsize\begin{eqnarray}
\Pi(q)&=&\sum^\infty_{j=a/2} (-)^{a/2-1} \int^1_0 dx \left\{ x(1-x) \right\}^{j-1}
\frac{(-Q^2)^{j-a/2}}{m^{2j}}\bigg[
-x(1-x)\Big\{ c_1 +a_2(1+j)
\nonumber\\&&
+\frac{a_3}{2!}(1+j)(2+j)+\cdots\,\,
+\frac{a_k}{(k-1)!}\prod^{k-1}_{l=1}(l+j)\Big\}+c_2 \bigg]
\end{eqnarray}}
holds only after making the following checks:
\begin{itemize}
\item $j=0$

{\footnotesize$$0=d_1+c_1+a_2+a_3+\cdots\,\,+a_k$$}

\item $j=1$
{\footnotesize\begin{eqnarray}
0=c_2+d_2+\int^1_0dx\left\{ -x(1-x) \right\}\{c_1+2a_2+3a_3+\cdots\,\,+ka_k\}
\nonumber\end{eqnarray}}

\item $j=2$
{\footnotesize\begin{eqnarray}
0=\int^1_0dx\left\{ -x(1-x) \right\}c_2 +\int^1_0dx \left\{ -x(1-x) \right\}^2
\left\{ c_1+\cdots\,\,+\frac{k(k+1)}{2}a_k \right\}
\nonumber\end{eqnarray}}

\item $j=3$
{\scriptsize\begin{eqnarray}
0=\int^1_0dx \left\{ -x(1-x) \right\}^2c_2 +\int^1_0dx \left\{ -x(1-x) \right\}^3
\left\{ c_1+\cdots\,\,+\frac{k(k+1)(k+2)}{3}a_k \right\}
\nonumber\end{eqnarray}}

\item $j=\frac{a}{2}-1$
{\footnotesize\begin{eqnarray}
&&0=\int^1_0dx \left\{ -x(1-x) \right\}^{\frac{a}{2}-2}c_2 \nonumber\\
&&+\int^1_0dx \left\{ -x(1-x) \right\}^{\frac{a}{2}-1}
\left\{ c_1+\cdots\,\,+\frac{k\cdots\,\,(k+\frac{a}{2}-3)(k+\frac{a}{2}-2)}{\frac{a}{2}-1}a_k \right\}
\nonumber\end{eqnarray}}
So we know the summation begins at $j=a/2$.

Substituting the values for the coefficients multiplying $J$'s, we have
explicitly checked that the above constraints are satisfied in our
calculation.
\end{itemize}
\end{appendix}


\end{document}